\documentclass[
 reprint, nofootinbib,
 amsmath,amssymb,
 aps, prd,
]{revtex4-2}
\usepackage[utf8]{inputenc}
\usepackage{graphicx}
\usepackage{dcolumn}
\usepackage{orcidlink}
\usepackage{subfigure}
\usepackage{footnote}
\usepackage[title]{appendix}
\usepackage{bm}
\usepackage{xcolor}
\usepackage{hyperref}
\begin{document}
\preprint{APS/123-QED}

\title{Maximal hypersurfaces and aspects of volume of the Kerr family of black holes}

\author{Suraj Maurya\orcidlink{0000-0001-6907-8584}}
 \email{p20200471@hyderabad.bits-pilani.ac.in (S. Maurya)}

\author{Sashideep Gutti\orcidlink{0000-0001-7555-8453}}%
 \email{sashideep@hyderabad.bits-pilani.ac.in (S. Gutti)}
 
 \author{Rahul Nigam\orcidlink{0000-0002-0497-5898}}%
 \email{rahul.nigam@hyderabad.bits-pilani.ac.in (R. Nigam)}
\affiliation{Birla Institute of Technology and Science Pilani (Hyderabad Campus), Hyderabad 500078, India}
\date{June 25, 2024}

\begin{abstract}
In Schwarzschild spacetime, Reinhart (1973) has shown the hypersurface $r_R = 3M/2$ (the subscript stands for ``Reinhart") to be a maximal hypersurface. This Reinhart radius $r_R$ plays a crucial role in evaluating the interior volume of a black hole. In this article, we find such a maximal hypersurface for the Kerr and Kerr-Newaman black holes. We obtain the analytical expression for the Reinhart radius as a function of the polar angle $\theta$ for a small $a/M$ limit for both the Kerr and Kerr-Newman black holes. We obtain the Reinhart radius using two independent methods: a) the vanishing trace of the extrinsic curvature and, b) the variational method.  We further use the Reinhart radius to obtain an analytical expression for the interior volume of the Kerr and Kerr-Newman black hole in the small $a/M$ limit and a generic charge $Q$. We define $\mathcal{\dot{V}}$ as the rate of change of the interior volume with respect to the ingoing null coordinate $v$ and then study its behavior under various scenarios, viz., particle accretion, the Penrose process, superradiance, and Hawking radiation. We show that while under the Penrose process and superradiance, the parameter $\mathcal{\dot{V}}$ increases just as the area of a black hole but under particle accretion, $\mathcal{\dot{V}}$ can have variable signs depending on the kinematical properties of the particle.  We further probe into the behavior of $\mathcal{\dot{V}}$ under Hawking radiation. These results provide important and very interesting clues toward the possible existence of laws governing the volume of black holes. 
 
\end{abstract}

\maketitle

\section{Introduction}  
 The recent astrophysical confirmation of the existence of black holes in our universe has brought the study of black holes and their properties to the forefront area of research. There has been a phenomenal amount of progress in our understanding of black holes over the past 50 years. The thermodynamic properties of black holes are one of the most intriguing aspects of a black hole. The entropy of a black hole has been successfully tied to the area of the black hole horizon. The black hole horizon is also shown to radiate thermally via Hawking radiation.\par
 
The laws of black hole mechanics dictate that under all classical processes (particle accretion, Penrose process, and superradiance), the area of the black hole always increases. The black hole loses its mass due to Hawking radiation and in this case, the area of a black hole decreases. These are now fairly well-understood phenomena. A relatively less understood concept is the interior volume of a black hole. This question has been addressed by various authors using different approaches. Parikh \cite{MP} discusses the definition of volume by constructing an invariant slice of the spacetime inside the black hole horizon. Cveti\ifmmode \check{c}\else \v{c}\fi{} \textit{et al}. \cite{Gibbons} have discussed the thermodynamical volume, $V_{th}$ inside a black hole in the presence of a varying cosmological constant $\Lambda$. More recently, Christodoulou and Rovelli \cite{CR} defined the black hole volume by choosing a spacelike hypersurface in the black hole interior. The spacelike hypersurface is obtained by setting up a variational problem using which one can obtain the hypersurface by maximizing the volume.  In this approach, the interior volume grows indefinitely as a function of the advance time. \par 

In the work of Christodoulou and Rovelli \cite{CR}, surprisingly, a crucial role is played by a spacelike hypersurface given by $r_R=3M/2$ (the subscript stands for ``Reinhart"), where $M$ is the ADM mass of the Schwarzschild black hole. This hypersurface was first discovered in 1973 by Reinhart \cite{Reinhart}. He showed that the surface $r_R=3M/2$ is a maximal hypersurface characterized by vanishing the trace of extrinsic curvature.  This Reinhart radius $r_{R}$ lies in the interior of the Schwarzschild black hole.  It is shown in \cite{CR, CL} that there is an intimate connection between maximal hypersurfaces and the interior volume of a black hole. It is proved in \cite{CR} that the interior volume of a black hole has a maximum contribution from the region near the Reinhart radius $r_{R}$. Therefore this region provides an excellent approximation for the interior volume of the black hole \cite{CR}. The volume of a Schwarzschild black hole is equal to $V=3\sqrt{3}\pi M^2v$, where $v$ is the advance time \cite{CR}.\par

The work along similar lines was extended to the Reissner–Nordstrom black hole in the paper \cite{YCO}  using the Reinhart radius of the Reissner-Norstrom black hole. The interior volume of the BTZ black hole is done in  \cite{SSR1}. The interior volume and its evolution due to the collapse of matter in D-dimensions is done in \cite{SRS}. The evolution of the interior volume of a black hole due to Hawking radiation is discussed in \cite{CL} where it is shown that the interior volume continues to increase even though the rate of increase of the volume decreases under Hawking radiation. A similar conclusion for the Reissner-Norstrom black hole is reached in \cite{YCO}.\par

Extending the analysis to astrophysically relevant rotating black holes, the interior volume estimate for the Kerr black hole is done in \cite{BJ}. Extensive work on the volume aspects of the Kerr-AdS black hole is done in \cite{XY}. In the works of \cite{BJ, XY}, they estimated the volume of the Kerr black hole using the volume maximization technique, where it assumes a hypersurface $r=const$ and then finds the radius $r$ that maximizes the volume. As highlighted in \cite{XY}, the $r=const$ hypersurface that was derived, is not a maximal hypersurface. \par

What is the equivalent of Reinhart radius $r_R=3M/2$ (maximal hypersurface) for Kerr spacetime? An analysis toward obtaining the maximal hypersurface in Kerr spacetime was done by Duncan \cite{Duncan}. He argued that the maximal hypersurface in the Kerr black hole is given by $r(\theta)$ where $\theta$ is the polar angle. Though the curve $r(\theta)$ was not found, Duncan evaluated an upper bound $r_{max}$ and a lower bound $r_{min}$,  within which the maximal hypersurface can be located. These bounds were used in studying aspects of holographic complexity and volume \cite{TJ}.\par  

The general expression for the maximal hypersurface in Kerr spacetime thus remains elusive. In our article, we obtain the exact expression for the Reinhart radius (maximal hypersurface) in the Kerr and the Kerr-Newman black holes for a small $a/M$ limit and a generic charge $Q$. We solve using two methods: the vanishing trace of extrinsic curvature and the maximization of volume technique. We show that the analytical expressions for the maximal hypersurface for slowly rotating black holes are surprisingly simple and given by $ r_R(\theta)= \frac{3M}{2} - \frac{a^2(14 - sin^2\theta)}{36M}$ for the Kerr black hole and Eq. (\ref{eqn77}) for the Kerr-Newman black hole. Using the expression for the Reinhart radius expressed as a function of $\theta$, we can now estimate the volume for the Kerr spacetime. We prove that for a slowly rotating Kerr black hole, the volume expression can be shown to be  $V^K = 3\sqrt{3}\pi M^2 v - \frac{16\sqrt{3}}{9}\pi a^2 v$. Similarly, the volume for the Kerr-Newman black hole is shown in Eq. (\ref{eqn90}). \par

The analytical closed-form expression for the volumes of Kerr and Kerr-Newman black holes lets one study the behavior of the volume under various dynamical processes concerning the black hole. We argue that since the volume is a linearly increasing function of late time parameter $v$, the rate of volume $\mathcal{\dot{V}}$ is physically more relevant. We show that under matter accretion, $\mathcal{\dot{V}}$ monotonically increases for both the Kerr and the Kerr-Newman black holes. We study the behavior of  $\mathcal{\dot{V}}$ under classical processes such as the Penrose process and superradiance where the mass and angular momentum of the black hole decrease. We show that the $\mathcal{\dot{V}}$ increases monotonically under both these processes. This behavior is similar to the area of a black hole. This strongly hints that the quantity defined by the rate of volume, $\mathcal{\dot{V}}$, might be governed by underlying laws just as the area of a black hole which monotonically increases under classical processes. We show that $\mathcal{\dot{V}}$ decreases under Hawking radiation for a Kerr black hole, indicating once again that $\mathcal{\dot{V}}$ copies the laws of the area of a black hole.\par 

In Sec. II, we examine the Kerr black hole and establish the metric element across various coordinate systems. In Sec. III, we use the traceless extrinsic curvature method to compute the Reinhart radius which gives the location of the maximal hypersurface inside the Kerr black hole. We verify the expression for the Reinhart radius for the Schwarzschild black hole by setting $a = 0$ and recalculating Duncan's result \cite{Duncan} for $r = const$ hypersurface. In Sec. IV, we use the variational method to set up the Euler-Lagrange equation to obtain the maximal hypersurface inside the Kerr black hole. In Sec. V, we compute the maximal interior volume of the Kerr black hole for the small $a/M$ limit. In Sec. VI, we review the Kerr-Newman black hole and establish the metric element in different coordinate systems. In Sec. VII, we use the traceless extrinsic method curvature to compute the Reinhart radius for the Kerr-Newman black hole. We verify the expression for the Reinhart radius for the Reisner-Nordstrom black hole by setting $a = 0$ and $r = const$. In Sec. VIII, we use the variational method to set up the Euler-Lagrange equation to obtain the maximal hypersurface inside the Kerr-Newman black hole. In Sec. IX, we calculate the maximal interior volume of the Kerr-Newman black hole for the small $a/M$ limit and a generic charge $Q$. In Sec. X, we give a clue about the volume law of black holes. In Sec. X A, we discuss the variation in the volume rate under the Penrose process, in Sec. X B, we discuss the variation in the volume rate under particle accretion, in Sec. X C, we discuss the variation in the volume rate under superradiance, and in Sec. X D, we discuss the variation in the volume rate under Hawking radiation for the Kerr black hole. In Sec. XI, we discuss the conclusion of the work. In Appendixes A and B, we discuss the determinant of the Kerr metric and solution of Eq. (\ref{eqn26}). In Appendix C, we discuss the determinant of the Kerr-Newman metric.

\section{The Kerr black hole}
 In this section, we describe a few useful relations of Kerr spacetime that will be used in the content that follows in the article. The Kerr metric in the Boyer-Lindquist coordinates $(t, r, \theta, \phi)$ is defined as
\begin{multline}\label{eqn1}
    ds^2 = - \frac{(\Delta - a^2sin^2{\theta})}{\rho^2}dt^2- \frac{4Mra sin^2{\theta}}{\rho^2} dtd\phi + \frac{\rho^2}{\Delta}dr^2\\ + \rho^2d{\theta^2} + \frac{Asin^2{\theta}}{\rho^2}d\phi^2 \\
    = g_{tt}dt^2 + 2g_{t\phi}dt d\phi + g_{rr}dr^2 + g_{\theta\theta}d\theta^2 + g_{\phi\phi}d\phi^2
\end{multline}
where the components of the metric (\ref{eqn1}) are defined as
\begin{equation}\label{eqn2}
    \begin{split}
        g_{tt} = - \frac{(\Delta - a^2sin^2{\theta})}{\rho^2} = -\bigg(1 - \frac{2Mr}{\rho^2}\bigg), \ \ g_{rr} = \frac{\rho^2}{\Delta},\\ \ \ g_{\theta\theta} = \rho^2, \ \ 
        g_{\phi\phi} = \frac{Asin^2{\theta}}{\rho^2}, \ \  g_{\phi t} = g_{t\phi} = - \frac{2Mra sin^2{\theta}}{\rho^2}
    \end{split}
\end{equation}
and the parameters, $\Delta, \rho^2, a,$ and $A$ are defined as
\begin{equation}\label{eqn3}
    \begin{split}
    \Delta = r^2 - 2Mr + a^2,\ \ \rho^2 = r^2 + a^2cos^2{\theta}\\ 
   a = J/Mc, \ \ A  = (r^2 + a^2)^2 - \Delta a^2sin^2{\theta}
    \end{split}
\end{equation}
Here $M$ and $J$ are the spacetime's ADM mass and angular momentum in the axially symmetric Kerr metric. Now, the determinant of the metric (\ref{eqn1}) is obtained as
\begin{equation}\label{eqn4}
       g^{(4)} = det(g_{\mu\nu}) = g_{rr}g_{\theta\theta}\big(g_{tt}g_{\phi\phi} - g^2_{t\phi}\big) = - \rho^4 sin^2{\theta}
\end{equation}
Defining the Kerr ingoing coordinates $(v, r, \theta, \tilde\phi)$ as
\begin{equation}\label{eqn5}
    \begin{split}
        v = t + \int \frac{r^2 + a^2}{\Delta}dr, \ \ \tilde\phi = \phi + \int \frac{a}{\Delta}dr\\
      \Rightarrow  dt = dv - \frac{r^2 + a^2}{\Delta}dr,\ \ d\phi = d\tilde\phi - \frac{a}{\Delta}dr
    \end{split}
\end{equation}
while the coordinates $(v, r, \theta, \tilde\phi)$ are well-behaved on the future horizon but singular on the past horizon. For example, a straightforward computation reveals that after a transformation to the Kerr ingoing coordinates $(v, r, \theta, \tilde\phi)$, the metric (\ref{eqn1}) becomes
\begin{multline}\label{eqn6}
    ds^2 = -\bigg(1 - \frac{2Mr}{\rho^2}\bigg)dv^2 + 2dvdr + \rho^2d\theta^2 - 2asin^2\theta drd\tilde\phi\\ - \frac{4Marsin^2\theta}{\rho^2}dvd\tilde\phi + \frac{Asin^2\theta}{\rho^2}d\tilde\phi^2
\end{multline}
These coordinates $(v, r, \theta, \tilde\phi)$ produce an extension of the Kerr metric across the future horizon. Now, we explore the Reinhart radius (a location of the maximal hypersurface) for the Kerr black hole in Sec. III
\section{Reinhart radius for the Kerr black hole}
When one considers Schwarzschild spacetime, there is an area radius given by $r_R=3M/2$ which we call the Reinhart radius \cite{Reinhart}, with a special property. The hypersurface given by $r_R=3M/2$ is a maximal hypersurface. This implies that the trace of extrinsic curvature vanishes. This hypersurface has an interesting application in estimating the interior volume of a black hole along the lines of the formalism developed in \cite{CR}. The hypersurface that generates the maximum volume is shown to get its main contribution near the radius $r_R=3M/2$. This lets one obtain an excellent approximation for the interior volume of the black hole. The Reinhart radius is similarly found for the BTZ black hole and plays a similar role in estimating the volume of the black hole \cite{SSR1}. \par
The question we now ask is, what is the hypersurface in the Kerr black hole? Duncan in \cite{Duncan} addressed the issue and has numerically evaluated and shown the maximal hypersurface graphically for various rotation parameters $a$. Though the analytical, closed-form expression for the maximal hypersurface has been elusive, Duncan has obtained a lower and upper bound where the maximal hypersurface can be located in the Kerr black hole.  Using $r=const$ hypersurfaces, he obtained the upper bound  $r_{max}$ and lower bound $r_{min}$ which we have shown in Eq. (\ref{eqn26'}), where the maximal hypersurface is located.\par

In this section, we obtain the exact analytical expressions for the maximal hypersurface for a slowly rotating Kerr black hole. We obtain the expression via two independent methods. In this section,  we obtain the expression for maximal hypersurface by calculating the trace of extrinsic curvature of the hypersurface and setting it to zero. The calculating trace of extrinsic curvature is equal to the divergence of the normal vector to the hypersurface. It is easy to see that the maximal hypersurface that is given by $r_R=3M/2$ for the Schwarzschild case will develop a dependence on the polar angle $\theta$. By symmetry considerations, it cannot depend on azimuthal angle $\phi$ and is time-independent. If we choose a section $\phi=const$, the intersection of the hypersurface with the $\phi=const$ is given by a curve in $(r,\theta)$ parameterized with $\lambda$ i. e.,  $r\equiv r(\lambda)$ and $\theta\equiv \theta(\lambda)$. The hypersurface is obtained by revolving the curve about the $z$-axis. The components of the tangent vector along a section of constant $\phi$ is 
\begin{equation}\label{eqn7}
    T^{r} = \frac{d r}{d\lambda}, \ \ T^{\theta} = \frac{d\theta}{d\lambda}
\end{equation}
It is easy to see that the coordinate $r$ varies monotonically as a function of $\theta$, where $0<\theta<\pi/2$.  If we take $\lambda = \theta$ as a parameter then $r(\lambda) \rightarrow r(\theta)$ and $\theta(\lambda)\rightarrow \theta$. Now, the tangent vector components become
\begin{equation}\label{eqn8}
    T^{r} = \frac{d r}{d \theta} = r_{,\theta}  \ and \ T^{\theta} = \frac{d \theta}{d \theta} = 1
\end{equation}
Using the above, we can deduce the components of the normal vector (contravariant) orthogonal to the above tangent vector are found to be
\begin{equation}\label{eqn9}
    n^{r} =  \frac{T^{\theta}}{g_{rr}} = \frac{\Delta}{\rho^2}, \ \ n^{\theta}  = -\frac{T^{r}}{g_{\theta\theta}} = -\frac{r_{,\theta}}{\rho^2}
\end{equation}
Upon  normalization of Eq. (\ref{eqn9}) we obtain
\begin{equation}\label{eqn10}
    N = \sqrt{g_{rr}(n^{r})^2 + g_{\theta\theta}(n^{\theta})^2} = \frac{1}{\rho}\sqrt{r^2_{,\theta} + \Delta} 
\end{equation}
Hence, normalized components of the normal vector i.e. $N^{\alpha} = (N^r,N^\theta)$ are define as
\begin{equation}\label{eqn11}
    N^{r} = \frac{n^r}{N} = \frac{\Delta}{\rho\sqrt{r^2_{,\theta} + \Delta}}, \ \ N^{\theta} = \frac{n^{\theta}}{N} = \frac{- r_{,\theta}}{\rho\sqrt{r^2_{,\theta} + \Delta}}
\end{equation}
The condition for the vanishing trace of extrinsic curvature is the same as making the normal vector divergenceless which is written as
\begin{widetext}
    \begin{equation}\label{eqn12}
    \begin{split}
    N^{\alpha}_{;\alpha} = \frac{1}{\sqrt{-g^{(4)}}}\frac{\partial}{\partial x^{\alpha}}(\sqrt{-g^{(4)}}N^{\alpha}) 
    =\frac{\partial}{\partial r}(\sqrt{-g^{(4)}}N^{r}) + \frac{\partial}{\partial \theta}(\sqrt{-g^{(4)}}N^{\theta}) = 0\\
   \Rightarrow \frac{\partial}{\partial r}\Bigg(\frac{\rho^2\Delta sin\theta}{\rho\sqrt{r^2_{,\theta} + \Delta}}\Bigg) - \frac{\partial}{\partial \theta}\Bigg(\frac{\rho^2 r_{,\theta}sin\theta }{\rho\sqrt{r^2_{,\theta} + \Delta}}\Bigg) 
   = \frac{\partial}{\partial r}\Bigg(\frac{\rho\Delta sin\theta}{\sqrt{r^2_{,\theta} + \Delta}}\Bigg) - \frac{\partial}{\partial \theta}\Bigg(\frac{\rho r_{,\theta} sin\theta }{\sqrt{r^2_{,\theta} + \Delta}}\Bigg) = 0
   \end{split}
\end{equation}
Substituting the value of $\rho$, $\Delta$, $g^{(4)}$, $N^r$ and $N^\theta$ in Eq. (\ref{eqn12}), we obtain the following partial differential equation
\begin{equation}\label{eqn13}
     \frac{\partial}{\partial r}\Bigg(\frac{\sin\theta\sqrt{r^2+a^2cos^2{\theta}}(r^2+a^2-2Mr)}{\sqrt{r^2_{,\theta} + r^2+a^2-2Mr}}\Bigg) - \frac{\partial}{\partial \theta}\Bigg(\frac{r_{,\theta} \sin\theta\sqrt{r^2+a^2cos^2{\theta}}}{\sqrt{r^2_{,\theta} + r^2+a^2-2Mr}}\Bigg) = 0
\end{equation}
We note here that the function $r,_{\theta}$ (a function of $\theta$ only) is to be determined from the above equation. After simplifying the above equation, we obtain the following expression
\begin{multline}\label{eqn14}
    (r^2_{,\theta} + r^2+a^2-2Mr)[sin\theta\{r(r^2+a^2-2Mr) + 2(r-M)(r^2+a^2cos^2{\theta})\} -(r_{,,\theta}sin\theta + r^2_{,\theta}cos\theta)(r^2+a^2cos^2{\theta})+ r_{,\theta}a^2sin^2\theta cos\theta]\\ - \sin\theta (r^2+a^2cos^2{\theta})[(r-M)(r^2+a^2-2Mr) -r^2_{,\theta}r_{,,\theta}]= 0
\end{multline}
After rearranging the expression in powers of $r$, Eq. (\ref{eqn14}) simplifies to the expression given below 
\begin{multline}\label{eqn15}
     (2sin\theta)r^5 - [7Msin\theta + sin\theta r_{,,\theta} + cos\theta r_{,\theta}]r^4 + [\{a^2(3+cos^2\theta) + 3(2M^2 + r^2_{,\theta})\}sin\theta + 2M(sin\theta r_{,,\theta} + cos\theta r_{,\theta})] r^3\\ - [ a^2\{(5 + 3cos^2\theta)Msin\theta + (1 + cos^2\theta) r_{,,\theta}sin\theta + 2cos^3\theta r_{,\theta}\} +(4Msin\theta + cos\theta r_{,\theta})r^2_{,\theta}] r^2 + a^2[sin\theta(r^2_{,\theta}+a^2)(1 + cos^2\theta) \\+ 2M(M+r_{,,\theta})sin\theta cos^2\theta ] r - a^2[r_{,\theta} (a^2+r^2_{,\theta}) cos\theta cos2\theta + \{2Mr^2_{,\theta} + (M + r_{,,\theta})a^2\}sin\theta cos^2\theta] = 0
\end{multline}
\end{widetext}
Now we can write Eq. (\ref{eqn15}) as follows
\begin{equation}\label{eqn16}
     a_1 r^5 - a_2 r^4 + a_3 r^3 - a_4 r^2 + a_5 r - a_6 = 0
\end{equation}
where the values of the coefficients $a_1, a_2, a_3, a_4, a_5,$ and $a_6$ are defined as
\begin{equation}\label{eqn17}
    a_1 =  2sin\theta
\end{equation}
\begin{equation}\label{eqn18}
    a_2 = 7Msin\theta + sin\theta r_{,,\theta} + cos\theta r_{,\theta}
\end{equation}
\begin{multline}\label{eqn19}
    a_3 = \{a^2(3+cos^2\theta) + 3(2M^2 + r^2_{,\theta})\}sin\theta \\+ 2M(sin\theta r_{,,\theta} + cos\theta r_{,\theta})
\end{multline}
\begin{multline}\label{eqn20}
    a_4 =  a^2[sin\theta\{(5 + 3cos^2\theta)M + (1 + cos^2\theta) r_{,,\theta}\} + 2cos^3\theta r_{,\theta}]\\+ (4Msin\theta + cos\theta r_{,\theta})r^2_{,\theta} 
 \end{multline}
 \begin{multline}\label{eqn21}
    a_5 = a^2sin\theta[(r^2_{,\theta}+a^2)(1 + cos^2\theta) + 2M(M+r_{,,\theta})cos^2\theta ]
 \end{multline}
 \begin{multline}\label{eqn22}
     a_6 =a^2[r_{,\theta} (a^2+r^2_{,\theta}) cos\theta cos2\theta \\+ \{2Mr^2_{,\theta} + (M + r_{,,\theta})a^2\}sin\theta cos^2\theta]
 \end{multline}
To check our analysis by comparing the results with the Schwarzschild black hole, we set the angular momentum per unit mass $a = 0$ and $r = const$; therefore, we get $r_{,\theta} = r_{,,\theta} = 0$. Now upon substituting these values in the above equation, we get $a_1 = 2sin\theta, a_2 = 7Msin\theta, a_3 = 6M^2sin\theta, a_4 = 0, a_5 = 0,$ and $a_6 =0$. Using these values in Eq. (\ref{eqn16}), we obtain the algebraic equation given below 
\begin{equation}\label{eqn23}
\begin{split}
    (2 r^5 - 7M r^4 + 6M^2 r^3)sin\theta = 0\\
    \Rightarrow r = 0, \ \ 2r^2 - 7M r + 6M^2  = 0
   \end{split} 
\end{equation}
Solving the above remaining quadratic equation in Eq. (\ref{eqn23}), we obtain the following result
\begin{equation}\label{eqn24}
    r = 0, \ \ r_R = \frac{3M}{2},  \ \ r_e = 2M
\end{equation}
The first root $r = 0$ is the singularity, the third root is the event horizon and the second root $r_R=3M/2$ is the Reinhart radius for the maximal hypersurface. The event horizon is a null surface and the Reinhart radius is a spacelike surface. Thus our equations reproduce the results known for Schwarzschild spacetime. Equation (\ref{eqn24}) exactly matches with the Schwarzschild black hole where $r_R$ is the Reinhart radius and $r_e$ is the event horizon of the black hole.\par
As a further check, we reproduce the upper and lower bounds obtained by Duncan \cite{Duncan}. Following Duncan, we set $r=const$ in Eq. (\ref{eqn14}), we find the roots for $\theta=0$ and $\theta=\pi/2$, and we get
\begin{equation}\label{eqn25'}
    (r^2 -2Mr + a^2)[2r^3 - 3Mr^2 + a^2(1+cos^2\theta)r - Ma^2cos^2\theta] = 0
\end{equation}
The solution of the second-order equation in the small bracket in Eq. (\ref{eqn25'}) gives the horizons $r_\pm$ of the Kerr black hole and solving the remaining third-order equation in the square bracket in Eq. (\ref{eqn25'}) at the boundaries $\theta = 0$ and $\theta = \pi/2$ gives the expression for $r_{min}$ and $r_{max}$. These expressions are given as follows
\begin{widetext}
    \begin{equation}\label{eqn26'}
       \begin{split}
            r_+ = M + \sqrt{M^2 - a^2},\ \ \ \ r_- = M - \sqrt{M^2 - a^2},\ \ \ \ r_{max}(\theta = \pi/2) = \frac{3M}{4} \Bigg(1 + \sqrt{1 - \frac{8a^2}{9M^2}}\Bigg),\\
            r_{min}(\theta = 0) = \frac{M}{2}\Bigg[1 + \Bigg\{1 + \sqrt{1 + \bigg(\frac{4a^2}{3M^2} - 1\bigg)^3}\Bigg\}^{1/3}- \bigg(\frac{4a^2}{3M^2} - 1\bigg)\Bigg\{1 + \sqrt{1 + \bigg(\frac{4a^2}{3M^2} - 1\bigg)^3}\Bigg\}^{-1/3}\Bigg]
       \end{split}
    \end{equation}
\end{widetext}
Hence, we have derived the results of the maximum and minimum radii\footnote{We have recalculated the value of $r$ at the boundaries i.e., $r_{max}$ at $\theta = \pi/2$ and $r_{min}$ at $\theta = 0$. We find that there is a typographical error in Duncan's expression of $r_{min}$. We have shown the correct value of $r_{min}$ in Eq. (\ref{eqn26'}). For the small $a/M$ limit,  we get the expressions for $r_{max} = \frac{3M}{2} - \frac{a^2}{3M}$ and $r_{min} = \frac{3M}{2} - \frac{4a^2}{9M}$.} of Duncan's paper \cite{Duncan}, which we have shown in Eq. (\ref{eqn26'}).\par
We now find the expression for the maximal hypersurface in the small $a/M$ limit. A solution for all $a$ might not be analytically feasible and might need numerical evaluation. We first deduce the few features possessed by the hypersurface that help in arriving at the solution. We note that the hypersurface should remain unaltered when we change the rotation parameter from $a$ to $-a$ respecting that the hypersurface should remain invariant under the change of spin of the black hole from clockwise to counterclockwise. The hypersurface should have reflection symmetry about $\theta=\pi/2$. This also implies that the slope of the curve $r(\theta)$ should vanish at $\theta=\pi/2$. The slope of the curve should similarly vanish at $\theta=0$ and $\theta=\pi$. If we set $a=0$ we should obtain the hypersurface $r_R=3M/2$. We now make an ansatz for the hypersurface based on the above considerations.
We suppose the general solution of Eq. (\ref{eqn16}) as a function $\theta$ is defined by
\begin{equation}\label{eqn25}
    r(\theta) = \frac{3M}{2} + a^2 g(\theta)
\end{equation}
where $r_{,\theta} = d r/ d\theta = a^2 g'(\theta)$ and $r_{,,\theta} = d^2 r/ d\theta^2 = a^2g''(\theta)$. We can neglect the terms with powers higher than $a^2$ in the $a/M$ limit. So, we get $a_6 = 0$ and Eq. (\ref{eqn16}) becomes 
\begin{equation}\label{eqn26}
    r[a_1r^4 -a_2 r^3 +a_3 r^2 - a_4 r +a_5] = 0
\end{equation}
Substitute these values of $r, r_{,\theta}, r_{,,\theta}, a_1, a_2, a_3, a_4,$ and $a_5$ in Eq. (\ref{eqn26}), and we obtain the following differential equation
\begin{equation}\label{eqn27}
   sin\theta g''(\theta) + cos\theta g'(\theta) - 2sin\theta g(\theta) = \frac{8}{9M}sin\theta - \frac{2}{9M}sin^3\theta
\end{equation}
We take the trial solution of the differential Eq. (\ref{eqn27}) as a series given below,
\begin{equation} \label{series}
     g(\theta) =  \sum_{n=0}^{\infty}c_n sin^n\theta 
\end{equation}
where $c_ns$ are undetermined constants. The above series implies
\begin{equation}\label{eqn28}
    \begin{split}
    g'(\theta) =\sum_{n=0}^{\infty} nc_n sin^{n-1}\theta cos\theta, \\  g''(\theta) = \sum_{n=0}^{\infty} [n(n-1)sin^{n-1}\theta - n^2sin^n\theta]c_n
    \end{split}
\end{equation}
Substitute the values of $g(\theta), g'(\theta),$ and $g''(\theta)$ from Eqs. (\ref{series}) and (\ref{eqn28}) into Eq. (\ref{eqn27}) and equate the constant term, coefficients of $sin\theta$ and $sin^3\theta$ on both sides, and we get
    \begin{equation}\label{eqn29}
       \begin{split}
            c_0 = -\frac{7}{18M},\ \ c_1 = 0,\ \ c_2 = \frac{1}{36M},\ \ c_3 = 0,\\ c_4 = k_1,\ \ c_{n+1} =\bigg[\frac{n^2-n+2}{(n+1)^2}\bigg]c_{n-1}; \ \ (n\ge 4)
       \end{split}
    \end{equation}
We note that $c_4= k_1$, where $k_1$ is an arbitrary constant. We note that a nonzero value of $k_1$ gives rise to an infinite series that goes to infinity at $\theta=\pi/2$ since $c_{n+1}$ for large $n$ is the same as $c_{n-1}$. The series therefore diverges. We obtain a solution where we obtain a finite value at $\theta=\pi/2$ when $k_1=0$. So we set the undetermined coefficient to $c_4$ to zero, thus obtaining a converging solution. Substituting these coefficients from Eq. (\ref{eqn29}) into Eq. (\ref{series}), we obtain a simple form given by     
    \begin{equation}\label{eqn30}
      g(\theta) = -\frac{1}{36M}(14 - sin^2\theta),
    \end{equation}
Hence, from Eqs. (\ref{eqn25}) and (\ref{eqn30}), we get
\begin{equation}\label{eqn31}
     r_R(\theta) \equiv  r(\theta) = \frac{3M}{2} + a^2g(\theta) = \frac{3M}{2} - \frac{a^2(14 - sin^2\theta)}{36M}
\end{equation}
where $r_R(\theta)$ is known as the Reinhart radius which is the general solution of Eq. (\ref{eqn26}). The Reinhart radius gives the location of the maximal hypersurface that maximizes the interior volume of the black hole. We rederive the maximal hypersurface using another method i.e., the variational method to find the maximal hypersurface in the Kerr black hole, which we discuss in Sec. IV. 
\section{A variational method for estimating the maximal hypersurface of the Kerr black hole}
We now obtain the maximal hypersurface (\ref{eqn31}) via an alternate method. This method is a generalization of the technique used by Christodoulou and Rovelli in \cite{CR} to estimate the hypersurface located in the interior of the black hole that gives a maximum contribution to the volume. They have considered the hypersurface of constant area radius $r=const$ and then estimated the $r$ for which there is maximal contribution. The maximization procedure yields $r_R=3M/2$ as the radius that dictates the interior volume of the black hole. Unsurprisingly, this coincides with the maximal hypersurface found by Reinhart \cite{Reinhart}. The method is adopted in the work by Bengtsson and Jakobsson \cite{BJ} for the Kerr black hole and \cite{XY} for the Kerr-AdS black hole. In \cite{BJ, XY}, they consider an $r=const$ hypersurface in Kerr and Kerr-AdS black holes respectively. They maximize the volume with respect to $r$ and arrive at the interior volume of the Kerr black hole. As highlighted in \cite{XY}, the hypersurface $r=const$ is not a maximal hypersurface in the Kerr family of black holes even though the hypersurface yields a good estimate for the interior volume of a black hole. In this section, we arrive at the maximal hypersurface using the volume maximization method used in \cite{CR}. Instead of using an $r=const$ hypersurface, we choose a hypersurface with $r$ as an unknown function of polar angle $\theta$ i.e., the hypersurface is given by $r -r(\theta)$=0. The maximization method used in \cite{BJ, XY}, yielded an algebraic equation when they considered $r=const$ hypersurface. In our approach since $r(\theta)$ is an unknown function that needs to be fixed via the maximization method, we hence obtain a variational problem. Again the expressions are untractable for a general $a/M$ but we show that it is possible to solve the problem in a closed form when we consider the small $a/M$ limit. This is therefore an alternate method of obtaining the maximal hypersurface. The details of the derivation are given below \par

The line element of the induced metric on the hypersurface $r-r(\theta)=0$ is $(dr=r,_\theta d\theta$, where $r,_{\theta}=dr/d\theta$)
\begin{multline}\label{eqn33}
    ds^2 = -\bigg(1 - \frac{2Mr}{\rho^2}\bigg)dv^2 + 2r_{,\theta}dvd\theta + \rho^2d\theta^2\\ - 2ar_{,\theta}sin^2\theta d\theta d\tilde\phi - \frac{4Marsin^2\theta}{\rho^2}dvd\tilde\phi + \frac{Asin^2\theta}{\rho^2}d\tilde\phi^2
    \\ 
    = g_{vv}dv^2 + 2g_{v\theta}dv d\theta + g_{\theta\theta}d\theta^2 + 2g_{\theta\tilde\phi}d\theta d\tilde\phi + 2g_{v\tilde\phi}dvd\tilde\phi \\+ g_{\tilde\phi\tilde\phi}d\tilde\phi^2
\end{multline}
where the components of the metric (\ref{eqn33}) are defined as
\begin{equation}\label{eqn34}
    \begin{split}
     g_{vv}= -\bigg(1 - \frac{2Mr}{\rho^2}\bigg), \ \ g_{v\theta} = g_{\theta v} = r_{,\theta},\ \  g_{\theta\theta} = \rho^2\\
     g_{\theta\tilde\phi} = g_{\tilde\phi\theta} = -ar_{,\theta}sin^2\theta,\ \  g_{\tilde\phi\tilde\phi}= \frac{Asin^2{\theta}}{\rho^2},\\ g_{v\tilde\phi} = g_{\tilde\phi v} = -\frac{2Mar}{\rho^2}sin^2{\theta}
     \end{split}
\end{equation}
The determinant of the metric (\ref{eqn33}) is defined as 
\begin{equation}\label{eqn35}
    g^{(3)} = det(g_{\mu\nu}) = - \rho^2(r^2_{,\theta} + \Delta)sin^2{\theta} 
\end{equation}
The volume bounded by the hypersurface $r - r(\theta) = 0$ is
\begin{equation}\label{eqn36}
    V^K(v) = \int_{V} \sqrt{-g^{(3)}}dvd\theta d\tilde\phi = 2\pi v\int sin\theta \sqrt{\rho^2(r^2_{,\theta} + \Delta)}d\theta
\end{equation}
where, the superscript $``K"$ stands for Kerr. This can be viewed as an extremization problem and our goal is to find the Euler-Lagrange equation. Substituting the value of $\rho^2$ and $\Delta$ from Eq. 
 (\ref{eqn3}) into Eq. (\ref{eqn36}), we get
  \begin{multline}\label{eqn37}
        L(r, r_{,\theta}, \theta) = sin\theta \sqrt{\rho^2(r^2_{,\theta} + \Delta)} \\= sin\theta \sqrt{(r^2 + a^2cos^2{\theta})(r^2_{,\theta} + r^2 - 2Mr + a^2)}
  \end{multline}
The Euler-Lagrange equation for the above Lagrangian is
\begin{equation}\label{eqn38}
    \frac{d}{d\theta}\bigg(\frac{\partial L}{\partial r_{,\theta}}\bigg) - \frac{\partial L}{\partial r} = 0 
\end{equation}
Now, substitute the expression for $r(\theta) = \frac{3M}{2} + a^2g(\theta)$ in Eq. (\ref{eqn38}), apply the small $a/M$ limit, keep the power up to $a^2$, and equate its coefficient to zero, and then we get
\begin{equation}\label{eqn39}
   sin\theta g''(\theta) + cos\theta g'(\theta) - 2sin\theta g(\theta) = \frac{8}{9M}sin\theta - \frac{2}{9M}sin^3\theta
\end{equation}
Eq. (\ref{eqn39}) is exactly matching with Eq. (\ref{eqn27}) whose solution is given in Eq. (\ref{eqn30}). 
\section{Volume and Volume rate of the Kerr black hole for a small value of \texorpdfstring{$\textbf{a/M}$}{Lg}}
The interior volume of the Kerr black hole is maximal when the area radius $r$ is reached at the Reinhart radius $r_R$. The expression of Reinhart radius for the Kerr black hole is given in Eq. (\ref{eqn31}). Therefore, at the radius $r = r_R$, the maximal volume of the Kerr black hole becomes 
\begin{widetext}
   \begin{equation}\label{eqn40}
      V^K(v) = \int_{V} \sqrt{-g^{(3)}}dvd\theta d\tilde\phi = 2\pi v\int_{0}^{\pi} sin\theta \sqrt{\rho^2(r^2_{,\theta} + \Delta)}d\theta = 2\pi v \int_{0}^{\pi} sin\theta \sqrt{(r^2_R + a^2cos^2\theta)(r^2_{R,\theta} + r^2_R + a^2 - 2Mr_R)}d\theta
   \end{equation}
Substitute the value of Reinhart radius $r_R$ from Eq. (\ref{eqn31}) into Eq. (\ref{eqn40}) and solve the integral for a small $a/M$ limit, and we get
\begin{multline}
  V^K(v)  = 2\pi v \int_{0}^{\pi} sin\theta \bigg[\bigg\{\bigg(\frac{3M}{2} - \frac{a^2(14 - sin^2\theta)}{36M}\bigg)^2 + a^2cos^2\theta\bigg\}\bigg\{\frac{a^4sin^2(2\theta)}{1296M^2} +\bigg(\frac{3M}{2} - \frac{a^2(14 - sin^2\theta)}{36M}\bigg)^2 + a^2\\ -2M\bigg(\frac{3M}{2} - \frac{a^2(14 - sin^2\theta)}{36M}\bigg)  \bigg\}\bigg]^{1/2}d\theta
\end{multline}
\end{widetext}
In the small $a/M$ limit the above integral gives the maximal volume of the Kerr black hole as
\begin{equation}\label{eqn41}
    V^K(v) = 3\sqrt{3}\pi M^2 v - \frac{16\sqrt{3}\pi a^2 v}{9} = V^{Sch}(v) - \frac{16\sqrt{3}\pi a^2 v}{9}
\end{equation}
where, the superscript $``Sch"$ stands for Schwarzschild. From Eq. (\ref{eqn41}), we can see that the maximal volume of the Kerr black hole is less than the maximal volume of the Schwarzschild black hole. The above statement has an intrinsic ambiguity since both volumes are proportional linearly to the ingoing coordinate $v$. There is no way to compare an ingoing coordinate $v$ between two separate spacetimes. It is, therefore, more meaningful to define a rate of volume as 
\begin{equation} \label{vdot}
    \mathcal{\dot{V}}=\frac{d V(v)}{dv}
\end{equation}
We can now compare the volume rate $\mathcal{\dot{V}}$ for Kerr and Schwarzschild black holes. We note from Eq. (\ref{eqn41}) that $\mathcal{\dot{V}}^K<\mathcal{\dot{V}}^{Sch}$ for the same mass of the black hole. We see in later sections that the quantity given by the volume rate $\mathcal{\dot{V}}$ has several interesting properties and its behavior is possibly governed by laws in a manner similar to the area of a black hole. 
\section{The Kerr-Newman black hole}
 In this section, we describe a few useful relations of Kerr-Newman spacetime that will be used in the content that follows in the article. The Kerr-Newman metric in the Boyer-Lindquist coordinates $(t, r, \theta, \phi)$ is defined as
\begin{multline}\label{eqn42}
    ds^2 = - \frac{(\Delta_Q - a^2sin^2{\theta})}{\rho^2}dt^2 - \frac{2a(2Mr-Q^2) sin^2{\theta}}{\rho^2}dtd\phi \\+ \frac{\rho^2}{\Delta_Q}dr^2 + \rho^2d{\theta^2} + \frac{A_Qsin^2{\theta}}{\rho^2}d\phi^2 \\
    = g_{tt}dt^2 + 2g_{t\phi}dt d\phi + g_{rr}dr^2 + g_{\theta\theta}d\theta^2 + g_{\phi\phi}d\phi^2
\end{multline}
where the components of the metric (\ref{eqn42}) are defined as
\begin{equation}\label{eqn43}
    \begin{split}
        g_{tt} = - \frac{(\Delta_Q - a^2sin^2{\theta})}{\rho^2}, \ \ g_{rr} = \frac{\rho^2}{\Delta_Q},\ \ g_{\theta\theta} = \rho^2\\
        g_{\phi\phi} = \frac{A_Qsin^2{\theta}}{\rho^2}, \ \  g_{\phi t} = g_{t\phi} = - \frac{a(2Mr -Q^2)sin^2{\theta}}{\rho^2}
    \end{split}
\end{equation}
and the parameters $\Delta_Q, \rho^2, a,$ and $A_Q$ are defined as
\begin{equation}\label{eqn44}
    \begin{split}
    \Delta_Q = r^2 - 2Mr + a^2 + Q^2,\ \ \rho^2 = r^2 + a^2cos^2{\theta}\\ 
   a = J/Mc, \ \ A_Q  = (r^2 + a^2)^2 - \Delta_Q a^2sin^2{\theta}
    \end{split}
\end{equation}
The determinant of the metric (\ref{eqn42}) is obtained as
\begin{equation}\label{eqn45}
   g^{(4)} = det(g_{\mu\nu}) = g_{rr}g_{\theta\theta}\big(g_{tt}g_{\phi\phi} - g^2_{t\phi}\big) = - \rho^4 sin^2{\theta}
\end{equation}
The metric (\ref{eqn42}) in the Kerr ingoing coordinates $(v, r, \theta, \tilde\phi)$ becomes
\begin{multline}\label{eqn46}
    ds^2 = -\bigg(1 - \frac{2Mr-Q^2}{\rho^2}\bigg)dv^2 + 2dvdr + \rho^2d\theta^2\\ - 2asin^2\theta drd\tilde\phi - \frac{2a(2Mr-Q^2)sin^2\theta}{\rho^2}dvd\tilde\phi \\+ \frac{A_Qsin^2\theta}{\rho^2}d\tilde\phi^2
\end{multline}
Now we explore the Reinhart radius (a location of the maximal hypersurface) for the Kerr-Newman black hole in Sec. VII.

\section{Reinhart radius for the Kerr-Newman black hole}
In the case of the Reisner-Nordstrom black hole, the location of the maximal hypersurface is denoted by $r_R$ which we call the Reinhart radius \cite{Reinhart, CR, SRS}, and is given by $r_R = (3M + \sqrt{9M^2 - 8Q^2)}/4$. The Reinhart radius $r_R$ is used to maximize the interior volume of the Reisner-Nordstrom black hole \cite{YCO}. The question we ask is, what is the Reinhart radius for the Kerr-Newman black hole? In this section, we obtain the exact analytical expression for the Reinhart radius of the Kerr-Newman black hole in the small $a/M$ limit and a generic charge $Q$. We obtained the Reinhart radius using the traceless extrinsic curvature method. The calculation of the trace of extrinsic curvature of the hypersurface is equal to the divergence of the normal vector to the hypersurface. As we know, the Kerr-Newman black hole is an axially symmetric black hole, so the Reinhart radius will depend on the polar angle $\theta$ and be independent of azimuthal angle $\phi$. If we choose a section $\phi = const$, the intersection of the hypersurface with angle $\phi$ is given by a curve $(r, \theta)$ parameterized with $\lambda$ i.e., $r\equiv r(\lambda)$ and $\theta\equiv \theta(\lambda)$. The hypersurface is obtained by revolving the curve about the $z$-axis. The components of the tangent vector along a section of constant $\phi$ are defined as
\begin{equation}\label{eqn47}
    T^{r} = \frac{d r}{d \lambda}, \ \ T^{\theta} = \frac{d\theta}{d\lambda}
\end{equation}
The coordinate $r$ varies monotonically as a function of $\theta$ where $0 < \theta < \pi/2$. If we take $\lambda = \theta$ as a parameter then $r(\lambda) \rightarrow r(\theta)$ and $\theta(\lambda)\rightarrow \theta$. Now, the tangent vector components become
\begin{equation}\label{eqn48}
    T^{r} = \frac{d r}{d\theta} = r_{,\theta}  \ and \ T^{\theta} = \frac{d\theta}{d \theta} = 1
\end{equation}
Similarly, we can deduce the contravariant components of the normal vector which are orthogonal to the above tangent vector and are found to be
\begin{equation}\label{eqn49}
    n^{r} =  \frac{T^{\theta}}{g_{rr}} = \frac{\Delta_Q}{\rho^2}, \ \ n^{\theta}  = -\frac{T^{r}}{g_{\theta\theta}} = -\frac{r_{,\theta}}{\rho^2}
\end{equation}
Upon normalization of Eq. (\ref{eqn49}), we obtained 
\begin{equation}\label{eqn50}
    N = \sqrt{g_{rr}(n^{r})^2 + g_{\theta\theta}(n^{\theta})^2} = \frac{1}{\rho}\sqrt{ r^2_{,\theta}+\Delta_Q} 
\end{equation}
Hence, normalized components of the normal vector i.e., $N^{\alpha} = (N^r,N^\theta)$ are defined as
\begin{equation}\label{eqn51}
    N^{r} = \frac{n^r}{N} = \frac{\Delta_Q}{\rho\sqrt{r^2_{,\theta}+\Delta_Q}}, \ \ N^{\theta} = \frac{n^{\theta}}{N} = \frac{- r_{,\theta}}{\rho\sqrt{r^2_{,\theta} +\Delta_Q}}
\end{equation}
Now we obtain the Reinhart radius $r_R$ by calculating the trace of extrinsic curvature and equating it to zero. The condition for the vanishing trace of extrinsic curvature is the same as making the normal vector divergenceless which is written as
\begin{widetext}
    \begin{equation}\label{eqn52}
    \begin{split}
    N^{\alpha}_{;\alpha} = \frac{1}{\sqrt{-g^{(4)}}}\frac{\partial}{\partial x^{\alpha}}(\sqrt{-g^{(4)}}N^{\alpha}) 
    =\frac{\partial}{\partial r}(\sqrt{-g^{(4)}}N^{r}) + \frac{\partial}{\partial \theta}(\sqrt{-g^{(4)}}N^{\theta}) = 0\\
   \Rightarrow \frac{\partial}{\partial r}\Bigg(\frac{\rho^2\Delta_Q sin\theta}{\rho\sqrt{r^2_{,\theta} +\Delta_Q}}\Bigg) - \frac{\partial}{\partial \theta}\Bigg(\frac{\rho^2 r_{,\theta}sin\theta }{\rho\sqrt{r^2_{,\theta} +\Delta_Q}}\Bigg) 
   = \frac{\partial}{\partial r}\Bigg(\frac{\rho\Delta_Q sin\theta}{\sqrt{r^2_{,\theta} +\Delta_Q}}\Bigg) - \frac{\partial}{\partial \theta}\Bigg(\frac{\rho r_{,\theta}sin\theta }{\sqrt{r^2_{,\theta} +\Delta_Q}}\Bigg) = 0
   \end{split}
\end{equation}
Substituting the value of $\rho$, $\Delta_Q$, $g^{(4)}$, $N^r$ and $N^\theta$ in Eq. (\ref{eqn52}), we get
\begin{equation}\label{eqn53}
     \frac{\partial}{\partial r}\Bigg(\frac{\sin\theta\sqrt{r^2+a^2cos^2{\theta}}(r^2+a^2-2Mr + Q^2)}{\sqrt{r^2_{,\theta} + r^2+a^2-2Mr + Q^2}}\Bigg) - \frac{\partial}{\partial \theta}\Bigg(\frac{r_{,\theta} \sin\theta\sqrt{r^2+a^2cos^2{\theta}}}{\sqrt{r^2_{,\theta} + r^2+a^2-2Mr + Q^2}}\Bigg) = 0
\end{equation}
We note here that the function $r,_{\theta}$ (a function of $\theta$ only) is to be determined from the above equation. After simplifying the above equation, we obtain the following expression
\begin{multline}\label{eqn54}
    (r^2_{,\theta} + r^2+a^2-2Mr + Q^2)[sin\theta\{r(r^2+a^2-2Mr + Q^2) + 2(r-M)(r^2+a^2cos^2{\theta})\} -(r_{,,\theta}sin\theta + r^2_{,\theta}cos\theta)(r^2+a^2cos^2{\theta})\\+ r_{,\theta}a^2sin^2\theta cos\theta)] - \sin\theta (r^2+a^2cos^2{\theta})[(r-M)(r^2+a^2-2Mr + Q^2) -r^2_{,\theta}r_{,,\theta}]= 0
\end{multline}
After rearranging the expression in powers of $r$, Eq. (\ref{eqn54}) simplifies to the expression given below 
\begin{multline}\label{eqn55}
     (2sin\theta)r^5 - [7Msin\theta + sin\theta r_{,,\theta} + cos\theta r_{,\theta}]r^4 + [\{a^2(3+cos^2\theta) + 3(2M^2 + r^2_{,\theta} + Q^2)\}sin\theta + 2M(sin\theta r_{,,\theta} + cos\theta r_{,\theta})] r^3\\ - [a^2[\{(5 + 3cos^2\theta)M + (1 + cos^2\theta) r_{,,\theta}\}sin\theta + 2cos^3\theta r_{,\theta}] +(4Msin\theta + cos\theta r_{,\theta})r^2_{,\theta} + Q^2\{(5M + r_{,,\theta})sin\theta + r_{,\theta}cos\theta\}] r^2\\ + [r^2_{,\theta}(2a^2cos^2\theta + a^2 + Q^2)sin\theta + Q^4sin\theta + a^2[sin\theta\{ (a^2(1+cos^2\theta) + Q^2(2+cos^2\theta) + 2M(M+r_{,,\theta}) cos^2\theta\}  + 2Mr_{,\theta} cos\theta cos2\theta]] r\\ - a^2[(a^2 + Q^2)\{(M + r_{,,\theta})cos^2\theta + r_{,\theta} cos\theta cos2\theta \}+ r^2_{,\theta}(2Msin\theta cos^2\theta + r_{,\theta} cos\theta cos2\theta)] = 0
\end{multline}
\end{widetext}
Now we can write Eq. (\ref{eqn55}) as follows
\begin{equation}\label{eqn56}
    b_1 r^5 - b_2 r^4 + b_3 r^3 - b_4 r^2 + b_5 r - b_6 = 0
\end{equation}
where the values of the coefficients $b_1, b_2, b_3, b_4, b_5,$ and $b_6$ are defined as
\begin{equation}\label{eqn57}
    b_1 =  2sin\theta
\end{equation}
\begin{equation}\label{eqn58}
    b_2 = 7Msin\theta + sin\theta r_{,,\theta} + cos\theta r_{,\theta}
\end{equation}
\begin{multline}\label{eqn59}
    b_3 = \{a^2(3+cos^2\theta) + 3(2M^2 + r^2_{,\theta} + Q^2)\}sin\theta \\+ 2M(sin\theta r_{,,\theta} + cos\theta r_{,\theta})
\end{multline}
\begin{multline}\label{eqn60}
    b_4 =   a^2[\{(5 + 3cos^2\theta)M + (1 + cos^2\theta) r_{,,\theta}\}sin\theta + 2cos^3\theta r_{,\theta}]\\ +(4Msin\theta + cos\theta r_{,\theta})r^2_{,\theta} 
    + Q^2\{(5M + r_{,,\theta})sin\theta + r_{,\theta}cos\theta\}
 \end{multline}
 \begin{multline}\label{eqn61}
    b_5 = r^2_{,\theta}(2a^2cos^2\theta + a^2 + Q^2)sin\theta + Q^4sin\theta\\ + a^2[sin\theta\{ (a^2(1+cos^2\theta) + Q^2(2+cos^2\theta) \\+ 2M(M+r_{,,\theta}) cos^2\theta\}  + 2Mr_{,\theta} cos\theta cos2\theta]
 \end{multline}
 \begin{multline}\label{eqn62}
     b_6 =a^2[(a^2 + Q^2)\{(M + r_{,,\theta}) cos^2\theta + r_{,\theta} cos\theta cos2\theta \}\\+ r^2_{,\theta}(2Msin\theta cos^2\theta + r_{,\theta} cos\theta cos2\theta)]
 \end{multline}
To check our analysis by comparing the results with the Reisner-Nordstrom black hole, we set the angular momentum per unit mass $a = 0$ and $r = const$; therefore, we get, $r_{,\theta} = r_{,,\theta} = 0$. Now upon substituting these values in the above equations, we get $b_1 = 2sin\theta, b_2 = 7Msin\theta, b_3 = (6M^2 + 3Q^2) sin\theta, b_4 = 5Q^2Msin\theta, b_5 = Q^4sin\theta,$ and $b_6 =0$. Now from Eq. (\ref{eqn56}), we get
\begin{equation}\label{eqn63}
    r[2r^4 - 7M r^3 + (6M^2 + 3Q^2)r^2 - 5Q^2Mr +Q^4]sin\theta = 0
\end{equation}
Equation (\ref{eqn63}) matches with the Reisner-Nordstrom black hole \cite{YCO}. Now solving the above remaining fourth-order equation in Eq. (\ref{eqn63}), we obtained the following results 
\begin{equation}\label{eqn64}
    r = 0, r^\pm_R = \frac{1}{4}(3M \pm \sqrt{9M^2 - 8Q^2}), r_\pm = M \pm \sqrt{M^2 - Q^2}
\end{equation}
The first root, i.e., $r = 0$, is the singularity; the second and third roots i.e., $r^{\pm}_R$, are the outer and inner Reinhart radii; and the fourth and fifth roots, i.e., $r_{\pm}$, are the outer and inner horizons of the black hole. The event horizon is a null surface and the Reinhart radius is a spacelike surface. Thus, Eq. (\ref{eqn64}) reproduces the results known for the Reisner-Nordstrom black hole. Here, the outer Reinhart radius $r^+_R$, which lies between the inner and outer horizons of the Reisner-Nordstrom black hole, is used to construct the maximal hypersurface.\par
We now find the expression of the Reinhart radius for the Kerr-Newman black hole in the small $a/M$ limit and a general charge $Q$. The solution for general $a$ might not be analytically feasible and might need a numerical solution. We first estimate some properties of the hypersurface that help us to reach the solution. The hypersurface should remain unaltered when we change the rotation parameter from $a$ to $-a$. The process to calculate the Reinhart radius of the Kerr-Newman black hole is the same as the Kerr black hole, but the result will differ due to the presence of charge $Q$. We suppose the general solution of Eq. (\ref{eqn56}) as a function $\theta$ which is defined as
\begin{equation}\label{eqn65}
    r(\theta) = \frac{1}{4}\big(3M + \sqrt{9M^2 - 8Q^2}\big) + a^2 g(\theta)
\end{equation}
where $r_{,\theta} = d r/ d\theta = a^2 g'(\theta)$ and $r_{,,\theta} = d^2 r/ d\theta^2 = a^2g''(\theta)$. We can neglect the terms with powers higher than $a^2$ (limit of small $a/M$) and substitute the values of $r, r_{,\theta}, r_{,,\theta}, b_1, b_2, b_3, b_4, b_5,$ and $b_6$ in Eq. (\ref{eqn56}), and we get
\begin{widetext}
\begin{multline}\label{eqn66}
    [27M^4 - 36M^2Q^2 + 8Q^4 + M(9M^2 - 8Q^2)^{3/2}]sin\theta g''(\theta) + [27M^4 - 36M^2Q^2 + 8Q^4 + M(9M^2 - 8Q^2)^{3/2}]cos\theta g'(\theta)\\ - 2[27M^4 - 42M^2Q^2 + 16Q^4 + M(9M^2 - 10Q^2)\sqrt{9M^2 - 8Q^2}]sin\theta g(\theta) - [21M^3 - 22MQ^2 + (7M^2 - 6Q^2)\sqrt{9M^2 - 8Q^2}]sin\theta\\ - [3M^3 -2MQ^2 + (M^2 - 2Q^2)\sqrt{9M^2 - 8Q^2}]sin\theta cos2\theta = 0\\
    \Rightarrow  [27M^4 - 36M^2Q^2 + 8Q^4 + M(9M^2 - 8Q^2)^{3/2}]sin\theta g''(\theta) + [27M^4 - 36M^2Q^2 + 8Q^4 + M(9M^2 - 8Q^2)^{3/2}]cos\theta g'(\theta)\\ - 2[27M^4 - 42M^2Q^2 + 16Q^4 + M(9M^2 - 10Q^2)\sqrt{9M^2 - 8Q^2}]sin\theta g(\theta) = 8[3M^3 - 3MQ^2 + (M^2 - Q^2)\sqrt{9M^2 - 8Q^2}]sin\theta\\ - 2[3M^3 -2MQ^2 + (M^2 - 2Q^2)\sqrt{9M^2 - 8Q^2}]sin^3\theta 
\end{multline}
\end{widetext}
Now, we can write Eq. (\ref{eqn66}) as follows
\begin{equation}\label{eqn67}
    sin\theta g''(\theta) + cos\theta g'(\theta) -\frac{2\beta}{\alpha}sin\theta g(\theta) = \frac{8\gamma}{\alpha}sin\theta - \frac{2\delta}{\alpha}sin^3\theta 
\end{equation}
where the parameters $\alpha, \beta, \gamma,$ and $\delta$  depend on the black hole's mass $M$ and charge $Q$ which are defined as
\begin{equation}\label{eqn68}
    \alpha = 27M^4 - 36M^2Q^2 + 8Q^4 + M\big(9M^2 - 8Q^2\big)^{3/2}
\end{equation}
\begin{multline}\label{eqn69}
     \beta = 27M^4 - 42M^2Q^2 + 16Q^4 + \big[ M\big(9M^2 - 10Q^2\big)\\ \times\sqrt{9M^2 - 8Q^2}\big]
\end{multline}
\begin{equation}\label{eqn70}
    \gamma = 3M^3 - 3MQ^2 + \big(M^2 - Q^2\big)\sqrt{9M^2 - 8Q^2}
\end{equation}
\begin{equation}\label{eqn71}
    \delta = 3M^3 -2MQ^2 + \big(M^2 - 2Q^2\big)\sqrt{9M^2 - 8Q^2}
\end{equation}
We take the trial solution of the differential Eq. (\ref{eqn67}) as a series given below
\begin{equation}\label{series1}
     g(\theta) =  \sum_{n=0}^{\infty}c_n sin^n\theta
\end{equation}
where $c_ns$ are undetermined constants. The above series implies
\begin{equation}\label{eqn72}
        \begin{split}
      g'(\theta) =\sum_{n=0}^{\infty} nc_n sin^{n-1}\theta cos\theta, \\
    g''(\theta) = \sum_{n=0}^{\infty} [n(n-1)sin^{n-1}\theta - n^2sin^n\theta]c_n
    \end{split}
\end{equation}
Substitute the values of $g(\theta), g'(\theta),$ and $g''(\theta)$ from Eqs. (\ref{series1}) and (\ref{eqn72}) into Eq. (\ref{eqn67}) and equate the constant term, coefficients of $sin\theta$ and $sin^3\theta$ on both sides, and we get 

\begin{equation}\label{eqn73}
    \begin{split}
        c_0 = -\bigg[\frac{4\gamma}{\beta} - \frac{2\alpha\delta}{(3\alpha+\beta)\beta}\bigg] = -\frac{p}{M},\ \  c_1 = 0,\\ c_2 = \frac{\delta}{3\alpha+\beta} = \frac{q}{M},\ \  c_3 = 0,\ \ c_4 = k_2,\\ c_{n+1} =\bigg[\frac{n^2-n+2}{(n+1)^2}\bigg]c_{n-1}; \ \ (n\ge 4)
        \end{split}
\end{equation}
As we know the parameters $\alpha, \beta, \gamma,$ and $\delta$ depend on the black hole's mass $M$ and charge $Q$, so the parameters $p$ and $q$ of Eq. (\ref{eqn73}), are also mass $M$ and charge $Q$ dependent. These parameters $p$ and $q$ are defined as
\begin{widetext}
\begin{equation}\label{eqn74}
    p = \frac{567 - 1458\big(\frac{Q}{M}\big)^2 + 1212\big(\frac{Q}{M}\big)^4 -320\big(\frac{Q}{M}\big)^6 + \big[189 - 402\big(\frac{Q}{M}\big)^2 + 244\big(\frac{Q}{M}\big)^4 - 32\big(\frac{Q}{M}\big)^6\big]\sqrt{9 - 8\big(\frac{Q}{M}\big)^2}}{1458-4293\big(\frac{Q}{M}\big)^2 + 4374\big(\frac{Q}{M}\big)^4 - 1700\big(\frac{Q}{M}\big)^6 + 160\big(\frac{Q}{M}\big)^8 + \big[486 - 1215\big(\frac{Q}{M}\big)^2 + 966\big(\frac{Q}{M}\big)^4 - 236\big(\frac{Q}{M}\big)^6\big]\sqrt{9 - 8\big(\frac{Q}{M}\big)^2}}
\end{equation}
\begin{equation}\label{eqn75}
   and \ \ q = \frac{3 -2\big(\frac{Q}{M}\big)^2 + \big[1 - 2\big(\frac{Q}{M}\big)^2\big]\sqrt{9 - 8\big(\frac{Q}{M}\big)^2}}{108 -150\big(\frac{Q}{M}\big)^2+40\big(\frac{Q}{M}\big)^4 +\big[36 -34\big(\frac{Q}{M}\big)^2\big]\sqrt{9 - 8\big(\frac{Q}{M}\big)^2}}
\end{equation}  
\end{widetext}
At $Q = 0$, we get $p = 7/18,\ \ c_0 = -7/18M$,\ \ $ q = 1/36$, and $c_2 = 1/36M$. These coefficients match with Eq. (\ref{eqn29}) of the Kerr black hole. Now, we note that from Eq. (\ref{eqn73}), $c_4= k_2$, where $k_2$ is an arbitrary constant. We note that a nonzero value of $k_2$ gives rise to an infinite series that goes to infinity at $\theta=\pi/2$ since $c_{n+1}$ for large $n$ is the same as $c_{n-1}$. The series therefore diverges. We choose a solution where we obtain a finite value at $\theta=\pi/2$ when $k_2 = 0$. Therefore we set the undetermined coefficient  $c_4$ to zero, thus obtaining a convergent solution. Substituting these coefficients from Eq. (\ref{eqn73}) into Eq. (\ref{series1}), we obtain a simple form of $g(\theta)$ given by
    \begin{equation}\label{eqn76}
        g(\theta) = -\frac{1}{M}\big(p - q sin^2\theta\big)
    \end{equation}
Hence from Eqs. (\ref{eqn65}) and (\ref{eqn76}), we get
    \begin{equation}\label{eqn77}
    \begin{split}
        r_R(\theta)  \equiv r(\theta) =\frac{1}{4}\big(3M + \sqrt{9M^2 - 8Q^2}\big) + a^2g(\theta)\\ = \frac{1}{4}\big(3M + \sqrt{9M^2 - 8Q^2}\big) - \frac{a^2}{M}\big(p - q sin^2\theta\big)
        \end{split}
    \end{equation}
where $r_R(\theta)$ is known as the Reinhart radius which is the solution of  Eq. (\ref{eqn56}).  We now use another method i.e., the variational method to find the maximal hypersurface in the Kerr-Newman black hole which we discuss in Sec. VIII.
\section{A variational method for estimating the maximal hypersurface of the Kerr-Newman black hole}
In this section, we use the variational approach to solve the Euler-Lagrange equation to obtain the maximal hypersurface. As we know from the analysis of the Kerr black hole, the Kerr-Newman black hole is an axially symmetric black hole; therefore, no hypersurface with $r = const$ is maximal. Instead of using $r = const$ hypersurface, we choose a hypersurface with $r$ as an unknown function of $\theta$ in the same way we had defined for the Kerr black hole i.e., $r - r(\theta) = 0$, so $dr = r_{,\theta} d\theta$, where $r_{,\theta} =  dr/d\theta$. The line element of the induced metric (\ref{eqn46}) on the hypersurface $r - r(\theta) = 0$ becomes
\begin{multline}\label{eqn78}
    ds^2 = -\bigg(1 - \frac{2Mr-Q^2}{\rho^2}\bigg)dv^2 + 2r_{,\theta}dvd\theta + \rho^2d\theta^2\\ - 2ar_{,\theta}sin^2\theta d\theta d\tilde\phi - \frac{2a(2Mr-Q^2)sin^2\theta}{\rho^2}dvd\tilde\phi \\+ \frac{A_Qsin^2\theta}{\rho^2}d\tilde\phi^2\\ 
    = g_{vv}dv^2 + 2g_{v\theta}dv d\theta + g_{\theta\theta}d\theta^2 + 2g_{\theta\tilde\phi}d\theta d\tilde\phi + 2g_{v\tilde\phi}dvd\tilde\phi \\+ g_{\tilde\phi\tilde\phi}d\tilde\phi^2
\end{multline}
where the components of the metric (\ref{eqn78}) are defined as
\begin{equation}\label{eqn79}
\begin{split}
     g_{vv}= -\bigg(1 - \frac{2Mr-Q^2}{\rho^2}\bigg), \ \ g_{v\theta} = g_{\theta v} = r_{,\theta},\ \  g_{\theta\theta} = \rho^2,\\
     g_{\theta\tilde\phi} = g_{\tilde\phi\theta} = -ar_{,\theta}sin^2\theta,\ \  g_{\tilde\phi\tilde\phi}= \frac{A_Qsin^2{\theta}}{\rho^2},\\ g_{v\tilde\phi} = g_{\tilde\phi v} = -\frac{a(2Mr-Q^2)}{\rho^2}sin^2{\theta}
     \end{split}
\end{equation}
The determinant of the metric (\ref{eqn78}) is defined as 
\begin{equation}\label{eqn80}
    g^{(3)} = det(g_{\mu\nu}) = - \rho^2(r^2_{,\theta} + \Delta_Q)sin^2{\theta} 
\end{equation}
The volume bounded by the hypersurface $r - r(\theta) = 0$ is
\begin{multline}\label{eqn81}
   V^{KN}(v) = \int_{V} \sqrt{-g^{(3)}}dvd\theta d\tilde\phi\\ = 2\pi v\int sin\theta \sqrt{\rho^2(r^2_{,\theta} + \Delta_Q)}d\theta
\end{multline}
where, the superscript $``KN"$ stands for Kerr-Newman. This can be viewed as an extremization problem and our goal is to find the Euler-Lagrange equation. Substituting the values of $\rho^2$ and $\Delta_Q$ from Eq. (\ref{eqn44}) into Eq. (\ref{eqn81}), we get
  \begin{multline}\label{eqn82}
        L(r, r_{,\theta}, \theta) = sin\theta \sqrt{\rho^2(r^2_{,\theta} + \Delta_Q)} \\= sin\theta \sqrt{(r^2 + a^2cos^2{\theta})(r^2_{,\theta} + r^2 - 2Mr + a^2 + Q^2)}
  \end{multline}
The Euler-Lagrange equation for the Lagrangian (\ref{eqn82}) is defined as
\begin{equation}\label{eqn83}
    \frac{d}{d\theta}\bigg(\frac{\partial L}{\partial r_{,\theta}}\bigg) - \frac{\partial L}{\partial r} = 0 
\end{equation}
Now, we substitute the expression for $r(\theta) = \frac{1}{4}\big(3M + \sqrt{9M^2 - 8Q^2}\big) + a^2g(\theta)$ in Eq. (\ref{eqn83}), apply the small $a/M$ limit, keep the power up to the $a^2$ term and equate its coefficient to zero, and we get,
\begin{equation}\label{eqn84}
    sin\theta g''(\theta) + cos\theta g'(\theta) -\frac{2\beta}{\alpha}sin\theta g(\theta) = \frac{8\gamma}{\alpha}sin\theta - \frac{2\delta}{\alpha}sin^3\theta 
\end{equation}
where the parameters $\alpha, \beta, \gamma,$ and $\delta$ are defined as
\begin{equation}\label{eqn85}
    \alpha = 27M^4 - 36M^2Q^2 + 8Q^4 + M\big(9M^2 - 8Q^2\big)^{3/2}
\end{equation}

\begin{multline}\label{eqn86}
     \beta = 27M^4 - 42M^2Q^2 + 16Q^4 + \big[ M\big(9M^2 - 10Q^2\big)\\ \times\sqrt{9M^2 - 8Q^2}\big]
\end{multline}

\begin{equation}\label{eqn87}
    \gamma = 3M^3 - 3MQ^2 + \big(M^2 - Q^2\big)\sqrt{9M^2 - 8Q^2}
\end{equation}

\begin{equation}\label{eqn88}
    \delta = 3M^3 -2MQ^2 + \big(M^2 - 2Q^2\big)\sqrt{9M^2 - 8Q^2}
\end{equation}
Equation (\ref{eqn84}) is exactly matching with Eq. (\ref{eqn67}) whose solution is given in Eq. (\ref{eqn76}). Now, we discuss the volume and volume rate of the Kerr-Newman black hole in Sec. IX.
\section{Volume and Volume rate of the Kerr-Newman black hole for a small value of \texorpdfstring{$\textbf{a/M}$}{Lg}}
In this section, we calculate the maximal volume and volume rate of the Kerr-Newman black hole for the small $a/M$ limit and a generic charge $Q$. The interior volume of the Kerr-Newman black hole is maximal when the area radius $r$ is reached at the Reinhart radius $r_R$. The expression of the Reinhart radius for the Kerr-Newman black hole is given in Eq. (\ref{eqn77}). Therefore at $r = r_R$, the maximal volume of the Kerr-Newman black hole is defined as
\begin{widetext}
   \begin{equation}\label{eqn89}
       V^{KN}(v) = 2\pi v\int_{0}^{\pi} sin\theta \sqrt{\rho^2(r^2_{,\theta} + \Delta_Q)}d\theta = 2\pi v \int_{0}^{\pi} sin\theta \sqrt{(r^2_R + a^2cos^2\theta)(r^2_{R,\theta} + r^2_R + a^2 - 2Mr_R + Q^2)}d\theta
   \end{equation}
Substitute the value of Reinhart radius $r_R$ for the Kerr-Newman black hole from Eq. (\ref{eqn77}) into Eq. (\ref{eqn89}) and solve the integral in a small $a/M$ limit and for a general charge $Q$, we get
\begin{multline}
    V^{KN}(v) = 2\pi v \int_{0}^{\pi} sin\theta \bigg[\bigg\{\frac{a^4q^2sin^2(2\theta)}{M^2} +\bigg(\frac{1}{4}\big(3M + \sqrt{9M^2 - 8Q^2}\big) - \frac{a^2}{M}\big(p - q sin^2\theta\big)\bigg)^2 + a^2\\ -2M\bigg(\frac{1}{4}\big(3M + \sqrt{9M^2 - 8Q^2}\big) - \frac{a^2}{M}\big(p - q sin^2\theta\big)\bigg) \bigg\}\bigg\{\bigg(\frac{1}{4}\big(3M + \sqrt{9M^2 - 8Q^2}\big) - \frac{a^2}{M}\big(p - q sin^2\theta\big)\bigg)^2 + a^2cos^2\theta\bigg\}\bigg]^{1/2}d\theta
\end{multline}
In the small $a/M$ limit the above integral gives the maximal volume of the Kerr-Newman black hole as follows  
\begin{multline}\label{eqn90}
    V^{KN}(v) = 4\pi v\bigg[\frac{1}{16}\big(3M + \sqrt{9M^2 - 8Q^2}\big)^2\bigg\{-Q^2 +\frac{M}{2}\big(3M + \sqrt{9M^2 - 8Q^2}\big) -\frac{1}{16}\big(3M + \sqrt{9M^2 - 8Q^2}\big)^2 \bigg\}\bigg]^{1/2}\\ - \pi v\Bigg[16a^2\bigg[Q^2M -\frac{1}{2}\big(3M + \sqrt{9M^2 - 8Q^2}\big)\bigg\{Q^2(3p - 2q) + \frac{M}{4}(6q-9p-2)\big(3M + \sqrt{9M^2 - 8Q^2}\big) + M^2\\+ \frac{1}{8}(3p-2q)\big(3M + \sqrt{9M^2 - 8Q^2}\big)^2\bigg\}\bigg]\bigg/6M\bigg[\big(3M + \sqrt{9M^2 - 8Q^2}\big)^2\bigg\{-Q^2 +\frac{M}{2}\big(3M + \sqrt{9M^2 - 8Q^2}\big) \\-\frac{1}{16}\big(3M + \sqrt{9M^2 - 8Q^2}\big)^2 \bigg\}\bigg]^{1/2}  \Bigg]
\end{multline}
The values of $p$ and $q$ are defined in Eqs. (\ref{eqn74}) and (\ref{eqn75}). Now, we can also define Eq. (\ref{eqn90}) as follows
\begin{multline}\label{eqn91}
    V^{KN}(v) = V^{RN}(v)  - \pi v\Bigg[16a^2\bigg[Q^2M -\frac{1}{2}\big(3M + \sqrt{9M^2 - 8Q^2}\big)\bigg\{Q^2(3p - 2q) + \frac{M}{4}(6q-9p-2)\big(3M + \sqrt{9M^2 - 8Q^2}\big) \\+ M^2 + \frac{1}{8}(3p-2q)\big(3M + \sqrt{9M^2 - 8Q^2}\big)^2\bigg\}\bigg]\bigg/6M\bigg[\big(3M + \sqrt{9M^2 - 8Q^2}\big)^2\bigg\{-Q^2 +\frac{M}{2}\big(3M + \sqrt{9M^2 - 8Q^2}\big) \\-\frac{1}{16}\big(3M + \sqrt{9M^2 - 8Q^2}\big)^2 \bigg\}\bigg]^{1/2}  \Bigg]
\end{multline}
\end{widetext}
where, the superscript $``RN"$ stands for Reisner-Nordstrom. From the above equation, we can see that the maximal volume of the Kerr-Newman black hole is less than the Reisner-Norsdrom black hole.  Similar to the discussion for the Kerr spacetime, we can define the volume rate for the Kerr-Newman black hole, given by Eq. (\ref{vdot}).  We can now compare the volume rate $\mathcal{\dot{V}}$ for the Kerr-Newman and Reisner-Nordstrom black holes. We note from Eq. (\ref{eqn91}) that $\mathcal{\dot{V}}^{KN}<\mathcal{\dot{V}}^{RN}$ for the same mass $M$ and charge $Q$ of the black hole.  
\section{Clues toward the possible existence of laws governing the volume of the black holes}
In the previous sections, we derived the expression for the maximal interior volume of the Kerr (\ref{eqn40}) and Kerr-Newman black holes (\ref{eqn90}). We note that the interior volume of the black holes was a linear function of ingoing coordinate $v$ that monotonically increases with $v$. The quantity concerning volume that displays more appropriately to study the behavior of volume is the volume rate $\mathcal{\dot{V}}$ as discussed in Eq. (\ref{vdot}). We now study the behavior of $\mathcal{\dot{V}}$ under various changes to the black hole. This study was first done in \cite{CL}, where they examined what happens to the volume of a black hole under Hawking radiation for a Schwarzschild black hole. They showed that the volume monotonically increases even under Hawking radiation. While this is true, we observe that in their paper, if we instead study the volume rate $\mathcal{\dot{V}}$, it can be inferred that the $\mathcal{\dot{V}}$ does indeed decrease under Hawking radiation. This is similar to the behavior of the area of the black hole since it is well known that the area of the black holes increases under particle accretion, the Penrose process, and superradiance but decreases under Hawking radiation. In this section, we show how the $\mathcal{\dot{V}}$ of the Kerr black hole behaves under these processes.

\subsection{Variation in the volume rate of the Kerr black hole under the Penrose process}
As we know, the Penrose process decreases the black hole's mass and angular momentum by an amount equal to (negative of) the energy and angular momentum of the infalling particle into the black hole \cite{Frolov, Paddy, Carrol}.  The decrease in the angular momentum is more than the decrease in the mass and the changes in the mass $(\delta M<0)$ and the angular momentum $(\delta J\ll 0)$ are related as 
\begin{equation}\label{eqn92}
    (\delta M - \Omega_H\delta J)>0
\end{equation}
where $\Omega_H$ is the angular velocity of the event horizon which is defined as
\begin{equation}\label{eqn93}
    \Omega_H = \frac{a}{r^2_+ + a^2} = \frac{J/M}{2M^2 + 2M\sqrt{M^2 - J^2/M^2}}
\end{equation}
where $r_+$ is the event horizon of the Kerr black hole. In the small angular momentum limit $J\ll M$, the above equation becomes
\begin{equation}\label{eqn94}
    \Omega_H =  \frac{a}{r^2_+ + a^2} \approx \frac{J}{4M^3}
\end{equation}
The maximal volume of the Kerr black hole is defined as
    \begin{equation}\label{eqn95}
          V^{K}(v) = 3\sqrt{3}\pi M^2 v - \frac{16\sqrt{3}\pi a^2 v}{9} 
    \end{equation}
The variation in the volume rate is defined as
\begin{equation}\label{eqn96}
    \delta \mathcal{\dot{V}}^K=\delta \bigg[\frac{\partial V^{K}(v)}{\partial v}\bigg] = \pi\sqrt{3}\bigg[\bigg(6M + \frac{32J^2}{9M^3}\bigg)\delta M - \frac{32J}{9M^2}\delta J\bigg]
\end{equation}
In the small angular momentum limit $J\ll M$, the above equation becomes
\begin{equation}\label{eqn97}
    \delta \mathcal{\dot{V}}^K = 6\sqrt{3}\pi M \bigg[\delta M - \frac{16}{27}\frac{J}{M^3}\delta J \bigg]
\end{equation}
From Eqs. (\ref{eqn94}) and (\ref{eqn97}), the variation in the volume rate becomes
\begin{equation}\label{eqn98}
     \delta \mathcal{\dot{V}}^K = 6\sqrt{3}\pi M\bigg[(\delta M - \Omega_H\delta J) - \frac{37}{27}\Omega_H\delta J\bigg]
\end{equation}
As we know, the change in the area $\delta A$ of the Kerr black hole is defined as
\begin{equation}\label{eqn99}
    \delta A = 8\pi\frac{a}{\Omega_H\sqrt{M^2-a^2}}(\delta M - \Omega_H \delta J)
\end{equation}
In the Penrose process $\delta A>0$ where $(\delta M - \Omega_H\delta J)>0$ and $\delta J< 0$, therefore from  Eq. (\ref{eqn98}), we get
\begin{equation}\label{eqn100}
    \delta \mathcal{\dot{V}}^K >0
\end{equation}
This shows that the variation in the volume rate always increases under the Penrose process.
\subsection{Variation in the volume rate of the Kerr black hole under superradiance}
As discovered by Penrose and Floyd \cite{RP}, it is possible to extract energy and angular momentum from the rotating black holes. In the superradiance, there is net energy that gets radiated to infinity when a scalar field propagates in the Kerr geometry. The energy flux $dE/dt$ and the amount of angular momentum $dJ/dt$ \cite{Paddy} (we note that $t$ is the Boyer-Lindquist time coordinate) that is falling into the horizon is defined as
\begin{equation}\label{eqn101}
    \frac{dE}{dt} = C_1\omega (\omega -m\Omega_H), \ \ \frac{dJ}{dt} = C_1 m (\omega -m\Omega_H)
\end{equation}
where $C_1$ is a constant and $\omega$ and $m$ are the frequency and angular momentum of the wave around the black hole spin axis. The complete descriptions of $C_1, \omega,$ and $m$ are given in \cite{Paddy}. Now, from Eq. (\ref{eqn101}), we get
\begin{equation}\label{eqn102}
    \frac{dE}{dt} - \Omega_H \frac{dJ}{dt} = C_1(\omega -m\Omega_H)^2 > 0
\end{equation}
As we know, the change in the black hole's mass is equivalent to rotational energy, i.e., $dM = dE$, so from Eq. (\ref{eqn99}), we can write the change in the black hole area as
\begin{equation}\label{eqn103}
    dA = 8\pi\frac{a}{\Omega_H\sqrt{M^2-a^2}}\bigg(\frac{dE}{dt} - \Omega_H \frac{dJ}{dt}\bigg)dt
\end{equation}
From  Eqs. (\ref{eqn102}) and  (\ref{eqn103}), we get
\begin{equation}\label{eqn104}
     dA = 8\pi\frac{a}{\Omega_H\sqrt{M^2-a^2}}C_1(\omega -m\Omega_H)^2dt > 0
\end{equation}
This seemingly agrees with the known fact that the area of the event horizon increases during any classical process. Similarly from Eq. (\ref{eqn98}), we can write the variation in the volume rate as follows
\begin{equation}\label{eqn105}
    d\mathcal{\dot{V}}^K = 6\sqrt{3}\pi M\bigg[\bigg(\frac{dE}{dt} - \Omega_H \frac{dJ}{dt}\bigg)dt - \frac{37}{27}\Omega_H\frac{dJ}{dt}dt\bigg]
\end{equation}
As we know, the energy and angular momentum radiate from the black hole, so $dE/dt <0$ and $dJ/dt<0$; therefore, from Eqs. (\ref{eqn102}) and (\ref{eqn105}), we get
\begin{equation}\label{eqn106}
    d\mathcal{\dot{V}}^K > 0
\end{equation}
This shows that the variation in the volume rate always increases under the superradiance phenomenon. 
\subsection{Variation in the volume rate of the Kerr black hole under an infalling particle}
Suppose a particle starts from a far faraway region and falls inside the event horizon of the Kerr black hole. The ingoing particle has positive energy and can have the angular momentum of either sign. But we know that under particle accretion, the area of the black hole increases. So we require that
\begin{equation}
     \delta M - \Omega_H\delta J > 0
\end{equation}
Unlike the Penrose process and superradiance, $\delta M>0$. If $\delta J$ is also positive, this implies there is an upper limit for $\delta J$, otherwise, there will be a violation of the area increase law. If we now examine the volume rate relation 
\begin{equation}\label{eqn98volrate}
     \delta \mathcal{\dot{V}}^K=6\sqrt{3}\pi M\bigg[(\delta M - \Omega_H\delta J) - \frac{37}{27}\Omega_H\delta J\bigg]
     \end{equation}
     Now the term inside the brackets $(\delta M - \Omega_H\delta J)$ is proportional to $\delta A$ and is always positive. Now if $\delta J<0$, it is easy to see that $\delta \mathcal{\dot{V}}$ is positive definite. If $\delta J$ is positive, then we see that the $\delta A$ term contributes a positive term while the $\delta J$ term contributes negatively. There could therefore be a $\delta J$ that can outweigh the positive contribution from the $\delta A $ term, making the volume rate $\mathcal{\dot{V}}$ negative. So it is possible to throw a particle with a large enough speed, with the same sense of rotation as the Kerr black hole, to make the volume rate decrease.
\subsection{Variation in the volume rate of the Kerr black hole under Hawking radiation}
     
During Hawking radiation, the area law is violated and the surface area of a black hole is expected to decrease ($\delta A <0$ and hence $\delta M-\Omega_H\delta J<0$). Moreover, the discussion in \cite{Frolov}, indicates that the black hole loses angular momentum faster than it loses mass. So we expect that $\delta J$ is negative for the case of Hawking radiation. The relation,
\begin{equation}\label{Hawkingrad}
     \delta \mathcal{\dot{V}}^K=6\sqrt{3}\pi M\bigg[(\delta M - \Omega_H\delta J) - \frac{37}{27}\Omega_H\delta J\bigg]
\end{equation}
gives that $\delta M - \Omega_H\delta J$ is negative but the term $-\frac{37}{27}\Omega_H\delta J$ is positive since $\delta J$ in Hawking radiation is shown to be negative \cite{Frolov}. So based on our study using small $J$ approximation, the outcome of $\delta \mathcal{\dot{V}}$ can be of either sign. This is by no means conclusive but the study is left for future considerations.

\section{Conclusions and Discussion}
In this article, we have obtained the closed-form expressions for the maximal hypersurface located between the inner and outer horizons of the Kerr and Kerr-Newman black holes for the small $a/M$ limit as shown in Eqs. (\ref{eqn31}) and (\ref{eqn77}).  This hypersurface is the generalization of the maximal hypersurface in Schwarzschild spacetime found by Reinhart in 1973 \cite{Reinhart}, to the Kerr family of black holes. Analogous to the volume of Schwarzschild spacetime,  the maximal hypersurfaces found for the Kerr black holes can be used to estimate the interior volume of the Kerr family of black holes.  Using the closed-form expressions for the maximal hypersurfaces, we have evaluated the closed-form expressions for the interior volume of the Kerr and Kerr-Newman black holes in the small $a/M$ limit as shown in Eqs. (\ref{eqn41}) and (\ref{eqn90}).\par
The volume for the Kerr black hole is estimated in terms of the ADM mass $M$ and angular momentum per unit mass $a$ (and a generic charge $Q$ for the Kerr-Newman black hole). The closed-form expression makes it convenient to study the behavior of the volume of the black hole under various processes such as absorption of a particle, the Penrose process, superradiance, and Hawking radiation using the variation of $M$ and $a$ in each process. An interesting question naturally arises. What happens to the volume of a black hole if the mass and angular momentum change due to these various processes?  Interestingly, the quantity that is sensitive to the processes is not the volume (which generally displays an increasing trend) but the rate of change of volume $\mathcal{\dot{V}}$ with respect to the ingoing null coordinate $v$. It is shown for the case of the Kerr black hole that under the Penrose process, the quantity $\mathcal{\dot{V}}$ monotonically increases. During superradiance, too, the quantity $\mathcal{\dot{V}}$ shows an increasing trend. It is shown in the article that if the black hole absorbs a particle, depending on the sign and angular momentum of the particle, $\mathcal{\dot{V}}$ can be negative or positive. A similar study of the behavior of volume change due to Hawking radiation is inconclusive regarding the trend of $\mathcal{\dot{V}}$. \par
We note that the study is done in the small $a/M$ limit. What is the behavior of $\mathcal{\dot{V}}$ when we relax the assumption and consider a more general $a/M$? This may be possible using a numerical study and is left for future considerations. Do the same trends continue in generic situations?  The analysis of variation of $\mathcal{\dot{V}}$ can be explored in a more generic case involving charge $Q$.  The trends of $\mathcal{\dot{V}}$ for the Kerr-Newman black hole are more subtle and are left for future considerations. \par

What is the significance of $\mathcal{\dot{V}}$?  The area of the black hole is intimately connected to the entropy of the black hole. The fact is that $\mathcal{\dot{V}}$ has a definite behavioral trend under various processes. Is this indicative of the possibility of associating $\mathcal{\dot{V}}$ with a thermodynamic interpretation? The behavior of $\mathcal{\dot{V}}$ hints toward the possible existence of laws concerning the volume of black holes. A thorough investigation, involving both analytical and numerical, might be needed to answer these questions and is left for future consideration. 
\begin{acknowledgments}
    We thank our institute BITS Pilani Hyderabad campus for providing the required infrastructure for this research work. S. M. thanks the funding agency, Council of Scientific and Industrial Research (CSIR), Government of India, File No. 09/1026(11329)/2021-EMR-I, for providing the necessary fellowship to support this research work.
\end{acknowledgments}
\appendix
\begin{widetext}
\section{Determinant of the Kerr metric}
The Kerr metric in the Boyer-Lindquist coordinates $(t, r, \theta, \phi)$ is defined as

\begin{equation}\label{A5}
    \begin{split}
    ds^2 = - \frac{(\Delta - a^2sin^2{\theta})}{\rho^2}dt^2- \frac{2Mra sin^2{\theta}}{\rho^2} (dtd\phi + d\phi dt)+ \frac{\rho^2}{\Delta}dr^2 + \rho^2d{\theta^2} + \frac{Asin^2{\theta}}{\rho^2}d\phi^2\\
    = g_{tt}dt^2 + g_{t\phi}dt d\phi + g_{\phi t}d\phi dt + g_{rr}dr^2 + g_{\theta\theta}d\theta^2 + g_{\phi\phi}d\phi^2
    \end{split}
\end{equation}
where the components of the metric (\ref{A5}) are defined as

\begin{equation}\label{A6}
    \begin{split}
        g_{tt} = - \frac{(\Delta - a^2sin^2{\theta})}{\rho^2}, \ \ g_{rr} = \frac{\rho^2}{\Delta},\ \ g_{\theta\theta} = \rho^2, \ \ 
        g_{\phi\phi} = \frac{Asin^2{\theta}}{\rho^2}, \ \  g_{\phi t} = g_{t\phi} = - \frac{2Mra sin^2{\theta}}{\rho^2}
    \end{split}
\end{equation}
and the parameters $\Delta, \rho^2, a,$ and $A$ are defined as
\begin{equation}\label{A7}
    \begin{split}
    \Delta = r^2 - 2Mr + a^2,\ \ \rho^2 = r^2 + a^2cos^2{\theta}, \ \ 
   a = J/Mc, \ \ A  = (r^2 + a^2)^2 - \Delta a^2sin^2{\theta}
    \end{split}
\end{equation}
From Eqs. (\ref{A6}) and (\ref{A7}), we can write the metric components $g_{tt}$ and $g_{\phi\phi}$ as
\begin{equation}\label{A2}
    \begin{split}
     g_{tt}= -\frac{(\Delta-a^2sin^2{\theta})}{\rho^2}= -\frac{(r^2 - 2Mr + a^2 - a^2sin^2{\theta})}{\rho^2} = -\frac{(r^2 - 2Mr + a^2cos^2{\theta})}{\rho^2} = -\frac{(\rho^2 - 2Mr)}{\rho^2} = -\bigg(1-\frac{2Mr}{\rho^2}\bigg),\\
      g_{\phi\phi} = \frac{Asin^2{\theta}}{\rho^2} =  \frac{[(r^2 + a^2)^2 - (r^2 - 2Mr + a^2) a^2sin^2{\theta}]sin^2{\theta}}{\rho^2} = \frac{[(r^2 + a^2)^2 - (r^2 + a^2) a^2sin^2{\theta} + 2Mra^2sin^2\theta]sin^2{\theta}}{\rho^2}\\
        =  \frac{[(r^2 + a^2)\{r^2 + a^2cos^2{\theta}\} + 2Mra^2sin^2\theta]sin^2{\theta}}{\rho^2} = \bigg[r^2 + a^2 + \frac{2Mra^2sin^2\theta}{\rho^2}\bigg]sin^2{\theta}
     \end{split}
\end{equation}
Equation (\ref{A5}) can be written in the matrix form as
    \begin{equation}\label{A9}
     g_{\mu\nu} = \begin{pmatrix}
    g_{tt} & g_{tr} & g_{t\theta} & g_{t\phi}\\
    g_{rt} & g_{rr} & g_{r\theta} & g_{r\phi}\\
    g_{\theta t}&g_{\theta r} & g_{\theta\theta} & g_{\theta\phi}\\
    g_{\phi t} & g_{\phi r} & g_{\phi\theta} & g_{\phi\phi}
  \end{pmatrix}_{4\times4}
  = \begin{pmatrix}
    g_{tt} & 0 & 0 & g_{t\phi}\\
    0 & g_{rr} & 0 & 0\\
    0 & 0 & g_{\theta\theta} & 0\\
    g_{\phi t} & 0 & 0 & g_{\phi\phi}\\
  \end{pmatrix}_{4\times4}
  \end{equation}
  The determinant of matrix (\ref{A9}) is
\begin{multline}\label{A10}
g^{(4)} = g_{rr}g_{\theta\theta}[g_{tt}g_{\phi\phi} - g^2_{t\phi}] = \frac{\rho^2}{\Delta}\times\rho^2\bigg[-\bigg(1-\frac{2Mr}{\rho^2}\bigg)\times\bigg(r^2 + a^2 + \frac{2Mra^2sin^2\theta}{\rho^2}\bigg)sin^2{\theta} - \frac{4M^2r^2a^2 sin^4{\theta}}{\rho^4}\bigg]\\
  =  \frac{\rho^4}{\Delta}\bigg[-\bigg(r^2 + a^2 + \frac{2Mra^2sin^2\theta}{\rho^2}\bigg)sin^2{\theta} + (r^2 +a^2)\frac{2Mr}{\rho^2}sin^2\theta + \frac{4M^2r^2a^2 sin^4{\theta}}{\rho^4} - \frac{4M^2r^2a^2 sin^4{\theta}}{\rho^4}\bigg]\\
 = \frac{\rho^4}{\Delta}\bigg[-\bigg(r^2 + a^2 + \frac{2Mra^2sin^2\theta}{\rho^2}\bigg)sin^2{\theta} + (r^2 +a^2)\frac{2Mr}{\rho^2}sin^2\theta\bigg] = \frac{\rho^4sin^2\theta}{\Delta}\bigg[-(r^2 + a^2) + \frac{2Mr}{\rho^2} [-a^2sin^2\theta + r^2 +a^2]\bigg] \\
= \frac{\rho^4}{\Delta}\bigg[-(r^2 + a^2) + \frac{2Mr}{\rho^2} [r^2 + a^2cos^2\theta ]\bigg]sin^2\theta = \frac{\rho^4}{\Delta}\bigg[-(r^2 + a^2) + \frac{2Mr}{\rho^2}\times \rho^2 \bigg]sin^2\theta = -\frac{\rho^4}{\Delta}\times \Delta sin^2\theta = - \rho^4sin^2\theta
\end{multline}
\section{Solution of Eq. (\ref{eqn26}) for the small a/M limit}
Let us calculate the terms $r, r^2, r^3,$ and $r^4$ in a small $a/M$ regime using the binomial expansion, as follows
\begin{multline}\label{B1}
        [r(\theta)] = \bigg[\frac{3M}{2} + a^2g(\theta)\bigg] = \bigg(\frac{3M}{2}\bigg)\bigg[1 + a^2g(\theta)\bigg(\frac{2}{3M}\bigg)\bigg],\ \  [r(\theta)]^2 = \bigg[\frac{3M}{2} + a^2g(\theta)\bigg]^2 = \bigg(\frac{3M}{2}\bigg)^2\bigg[1 + 2a^2g(\theta)\bigg(\frac{2}{3M}\bigg)\bigg],\\
         [r(\theta)]^3 = \bigg[\frac{3M}{2} + a^2g(\theta)\bigg]^3 = \bigg(\frac{3M}{2}\bigg)^3\bigg[1 + 3a^2g(\theta)\bigg(\frac{2}{3M}\bigg)\bigg],\ \ 
        [r(\theta)]^4 = \bigg[\frac{3M}{2} + a^2g(\theta)\bigg]^4 = \bigg(\frac{3M}{2}\bigg)^4\bigg[1 + 4a^2g(\theta)\bigg(\frac{2}{3M}\bigg)\bigg]
\end{multline}
Now, the parameters $a_1, a_2, a_3, a_4, a_5,$ and $a_6$ can be approximated in the small $a/M$ limit and keep the powers up to $a^2$ terms, where we get
    \begin{multline}\label{B2}
        a_1 = 2sin\theta, \ \ a_2 =  7Msin\theta + sin\theta r_{,,\theta} + cos\theta r_{,\theta} =  7Msin\theta + a^2[sin\theta g''(\theta) + cos\theta g'(\theta)],\\
       a_3 =[a^2(3+cos^2\theta) + 3(2M^2 + r^2_{,\theta})]sin\theta + 2M(sin\theta r_{,,\theta} + cos\theta r_{,\theta}) = a^2[(3 + 4cos^2\theta)sin\theta + 2M\{sin\theta g''(\theta) + cos\theta g'(\theta)\}]\\ + 6M^2sin\theta,\\
       a_4 = a^2[M(5+3cos^2\theta)sin\theta + (1+cos^2\theta)(sin\theta r_{,,\theta} + cos\theta r_{,\theta}) + 2cos^3\theta r_{,\theta}] + (4Msin\theta + cos\theta r_{,\theta})r^2_{,\theta} = [M(5+3cos^2\theta)sin\theta] a^2,\\
       a_5 =a^2sin\theta[(r^2_{,\theta}+a^2)(1 + cos^2\theta) + 2M(M+r_{,,\theta})cos^2\theta ]= [2M^2sin\theta cos^2\theta] a^2,\\ 
       a_6 = a^2[r_{,\theta} (a^2+r^2_{,\theta}) cos\theta cos2\theta + \{2Mr^2_{,\theta} + (M + r_{,,\theta})a^2\}sin\theta cos^2\theta]  = 0
\end{multline}
Now, multiplying $a_1, a_2, a_3,$ and $a_4$ with $r^4, r^3, r^2,$ and $r$, we get
\begin{multline}\label{B3}
    a_1 r^4 = 2sin\theta \times \bigg[\bigg(\frac{3M}{2}\bigg)^4 + 4g\bigg(\frac{3M}{2}\bigg)^3 a^2\bigg] = 2sin\theta \times\bigg(\frac{3M}{2}\bigg)^4 + \bigg[8g(\theta)sin\theta \times\bigg(\frac{3M}{2}\bigg)^3\bigg] a^2,\\
    a_2 r^3 = 7M\times \bigg(\frac{3M}{2}\bigg)^3sin\theta + \bigg[21Msin\theta\times\bigg(\frac{3M}{2}\bigg)^2g(\theta) + \bigg(\frac{3M}{2}\bigg)^3(sin\theta g''(\theta) + cos\theta g'(\theta)) \bigg]a^2,\\
    a_3 r^2 = \frac{27M^3sin\theta}{2} + \bigg(\frac{3M}{2}\bigg)^2[(3 + cos^2\theta)sin\theta + 2M\{sin\theta g''(\theta) + cos\theta g'(\theta)\}] + 18M^3sin\theta g(\theta)]a^2,\\
    a_4 r = \bigg[\frac{3M^2}{2}(5 + 3cos^2\theta)sin\theta \bigg] a^2
\end{multline}
Now, substitute the values of $a_1 r^4, a_2 r^3, a_3 r^2, a_4 r,$ and $a_5$ into Eq. (\ref{eqn26}) and equate the coefficients of $a^2$ to zero, we get
\begin{multline}\label{B4}
    g''(\theta)\bigg[-\bigg(\frac{3M}{2}\bigg)^3sin\theta + \bigg(\frac{3M}{2}\bigg)^2\times 2Msin\theta \bigg] + g'(\theta)\bigg[-\bigg(\frac{3M}{2}\bigg)^3cos\theta + \bigg(\frac{3M}{2}\bigg)^2\times 2Mcos\theta \bigg]\\+ g(\theta)\bigg[8sin\theta\times \bigg(\frac{3M}{2}\bigg)^3 - 21M\times \bigg(\frac{3M}{2}\bigg)^2sin\theta + 18M^3\bigg] + \bigg[\bigg(\frac{3M}{2}\bigg)^2(3+cos^2\theta) - \bigg(\frac{3M^2}{2}\bigg)(5 + 3cos^2\theta)+2M^2 cos^2\theta\bigg]sin\theta = 0\\
    \Rightarrow \frac{9}{8}M^3sin\theta g''(\theta) + \frac{9}{8}M^3cos\theta g'(\theta) - \frac{9}{4}M^3sin\theta g(\theta) - \frac{M^2}{4}[(3 + cos^2\theta)sin\theta] = 0\\
    \Rightarrow sin\theta g''(\theta) + cos\theta g'(\theta) - 2sin\theta g(\theta) = \frac{8}{9M}sin\theta - \frac{2}{9M}sin^3\theta
\end{multline}
\section{Determinant of the Kerr-Newmann metric}
The Kerr-Newman metric in the Boyer-Lindquist coordinates $(t, r, \theta, \phi)$ is defined as
\begin{multline}\label{C5}
    ds^2 = - \frac{(\Delta_Q - a^2sin^2{\theta})}{\rho^2}dt^2 + \frac{\rho^2}{\Delta_Q}dr^2 +  \rho^2d{\theta^2}
     - \frac{a(2Mr - Q^2)sin^2{\theta}}{\rho^2} (dtd\phi + d\phi dt)+ \frac{A_Qsin^2{\theta}}{\rho^2}d\phi^2
    = g_{tt}dt^2 +g_{rr}dr^2 \\+ g_{\theta\theta}d\theta^2 +  g_{t\phi}dt d\phi + g_{\phi t}d\phi dt +  g_{\phi\phi}d\phi^2
\end{multline}
where the components of the metric (\ref{C5}) are defined as
\begin{equation}\label{C6}
    \begin{split}
        g_{tt} = - \frac{(\Delta_Q - a^2sin^2{\theta})}{\rho^2}, \ \ g_{rr} = \frac{\rho^2}{\Delta_Q},\ \ g_{\theta\theta} = \rho^2, \ \ 
        g_{\phi\phi} = \frac{A_Qsin^2{\theta}}{\rho^2}, \ \  g_{\phi t} = g_{t\phi} = - \frac{a(2Mr - Q^2) sin^2{\theta}}{\rho^2}
    \end{split}
\end{equation} 
and the parameters $\Delta_Q, \rho^2, a,$ and $A_Q$ are defined as
\begin{equation}\label{C7}
    \begin{split}
        \Delta_Q = r^2 - 2Mr + a^2 + Q^2 = \Delta + Q^2,\ \  a = J/Mc, 
        \rho^2 = r^2 + a^2cos^2{\theta}, \ \ A_Q  = (r^2 + a^2)^2 - \Delta_Q a^2sin^2{\theta}
    \end{split}
\end{equation}
The components $g_{tt}$ and $g_{\phi\phi}$ of the metric (\ref{C5}) can be written as
    \begin{multline}\label{C8}
        g_{tt} = - \frac{(\Delta_Q - a^2sin^2{\theta})}{\rho^2} = -\frac{(r^2 - 2Mr + a^2 - a^2sin^2{\theta}+ Q^2)}{\rho^2} = -\frac{(r^2 + a^2cos^2{\theta} - 2Mr + Q^2)}{\rho^2} = -\bigg(1-\frac{2Mr-Q^2}{\rho^2}\bigg)\\
        g_{\phi\phi} = \frac{A_Qsin^2{\theta}}{\rho^2} = \frac{[ (r^2 + a^2)^2 - \Delta_Q a^2sin^2{\theta}]sin^2{\theta}}{\rho^2} =  \frac{[(r^2 + a^2)^2 - (r^2 - 2Mr + a^2 + Q^2) a^2sin^2{\theta}]sin^2{\theta}}{\rho^2}\\
        =  \frac{[(r^2 + a^2)\{r^2 + a^2 - a^2sin^2{\theta}\} + (2Mr - Q^2)a^2sin^2\theta]sin^2{\theta}}{\rho^2} = \frac{[(r^2 + a^2)\rho^2 + (2Mr - Q^2)a^2sin^2\theta]sin^2{\theta}}{\rho^2}\\ = \bigg[r^2 + a^2 + \frac{(2Mr - Q^2) a^2sin^2\theta}{\rho^2}\bigg]sin^2{\theta}
    \end{multline} 
Equation (\ref{C5}) can be written in the matrix form as
    \begin{equation}\label{C9}
     g_{\mu\nu} = \begin{pmatrix}
    g_{tt} & g_{tr} & g_{t\theta} & g_{t\phi}\\
    g_{rt} & g_{rr} & g_{r\theta} & g_{r\phi}\\
    g_{\theta t}&g_{\theta r} & g_{\theta\theta} & g_{\theta\phi}\\
    g_{\phi t} & g_{\phi r} & g_{\phi\theta} & g_{\phi\phi}
  \end{pmatrix}_{4\times4}
  = \begin{pmatrix}
    g_{tt} & 0 & 0 & g_{t\phi}\\
    0 & g_{rr} & 0 & 0\\
    0 & 0 & g_{\theta\theta} & 0\\
    g_{\phi t} & 0 & 0 & g_{\phi\phi}\\
  \end{pmatrix}_{4\times4}
  \end{equation}
  The determinant of matrix (\ref{C9}) is
\begin{multline}\label{C10}
g^{(4)} = g_{rr}g_{\theta\theta}[g_{tt}g_{\phi\phi} - g^2_{t\phi}]\\ = \frac{\rho^2}{\Delta_Q}\times\rho^2\bigg[-\bigg(1-\frac{(2Mr-Q^2)}{\rho^2}\bigg)\times\bigg(r^2 + a^2 + \frac{(2Mr-Q^2)a^2sin^2\theta}{\rho^2}\bigg)sin^2{\theta} - \frac{a^2(2Mr - Q^2)^2 sin^4{\theta}}{\rho^4}\bigg]\\
  =  \frac{\rho^4sin^2\theta}{\Delta_Q}\bigg[-\bigg(r^2 + a^2 + \frac{(2Mr-Q^2)a^2sin^2\theta}{\rho^2}\bigg) + \frac{(r^2 +a^2)(2Mr-Q^2)}{\rho^2} + \frac{a^2(2Mr - Q^2)^2 sin^2{\theta}}{\rho^4} - \frac{a^2(2Mr - Q^2)^2 sin^2{\theta}}{\rho^4}\bigg]\\
 = -\frac{\rho^4sin^2\theta}{\Delta_Q}\bigg[\bigg(r^2 + a^2 - \frac{(2Mr-Q^2)(r^2 + a^2-a^2sin^2\theta)}{\rho^2}\bigg)\bigg]
= -\frac{\rho^4sin^2\theta}{\Delta_Q}\bigg[(r^2 + a^2) - \frac{(2Mr-Q^2)}{\rho^2} [r^2 + a^2cos^2\theta ]\bigg]\\ = -\frac{\rho^4sin^2\theta}{\Delta_Q}\bigg[(r^2 + a^2) - \frac{(2Mr-Q^2)}{\rho^2}\times \rho^2 \bigg]= -\frac{\rho^4sin^2\theta}{\Delta_Q}\bigg[r^2 + a^2 - 2Mr + Q^2 \bigg] = -\frac{\rho^4sin^2\theta}{\Delta_Q}\times \Delta_Q  = - \rho^4sin^2\theta
\end{multline}
  \end{widetext}
  
\bibliography{bibitems}

\end{document}